# Stable Periodic Orbits for Spacecrafts around Minor Celestial Bodies

Yu Jiang[1,2], Juergen Schmidt[3], Hengnian Li[1], Xiaodong Liu[3], Yue Yang[4]

1. State Key Laboratory of Astronautic Dynamics, Xi'an Satellite Control Center, Xi'an 710043, China
2. School of Aerospace Engineering, Tsinghua University, Beijing 100084, China
3. Astronomy Research Unit, University of Oulu, Finland
4. School of software engineering, Xi'an Jiao Tong University, Xi'an 710049, China

Y. Jiang (✉) e-mail: jiangyu_xian_china@163.com (corresponding author)

**Abstract**. We are interested in stable periodic orbits for spacecrafts in the gravitational field of minor celestial bodies. The stable periodic orbits around minor celestial bodies are useful not only for the mission design of the deep space exploration, but also for studying the long-time stability of small satellites in the large-size-ratio binary asteroids. The irregular shapes and gravitational fields of the minor celestial bodies are modeled by the polyhedral model. Using the topological classifications of periodic orbits and the grid search method, the stable periodic orbits can be calculated and the topological cases can be determined. Furthermore, we find five different types of stable periodic orbits around minor celestial bodies: A) Stable periodic orbits generated from the stable equilibrium points outside the minor celestial body; B) Stable periodic orbits continued from the unstable periodic orbits around the unstable equilibrium points; C) Retrograde and nearly circular periodic orbits with zero-inclination around minor celestial bodies; D) Resonance periodic orbits; E) Near-surface inclined periodic orbits. We take asteroid 243 Ida, 433 Eros, 6489 Golevka, 101955 Bennu, and the comet 1P/Halley for examples.

## 1. Introduction

Missions to minor celestial bodies (here including asteroids and comets) and the discovery of large-size-ratio binary asteroids and triple asteroids make the study of stable periodic orbits around minor celestial bodies important [1-10]. Several previous studies investigated the periodic orbits around minor celestial bodies, such as Scheeres et al. [11], Elipe and Lara [12], Palacián et al. [13], Vasilkova [14], Yu and Baoyin [15], Jiang et al. [16], Jiang and Baoyin [17], Ni et al. [18].

Hamilton and Burns [19] studied the orbital stability zones around asteroid by assuming the asteroid to be a point mass and considering the solar radiation. Scheeres et al. [11] used the expansion of spherical harmonics method to model the gravitational field of 433 Eros and computed direct, nearly circular, equatorial periodic orbits in the body-fixed frame of Eros. Elipe and Lara [12] used a rotating straight segment to model the irregular shape of asteroid 433 Eros, and calculated periodic orbits around the segment. Vasilkova [14] used a triaxial ellipsoid to model the elongated asteroid and calculated several periodic orbits around equilibrium points of the triaxial ellipsoid. Palacián et al. [13] investigated the invariant manifold, periodic orbits, and quasi-periodic orbits around a rotating straight segment. Wang et al. [20] used the perturbation expansion with 2-order Legendre spherical harmonic coefficients to model the gravitational field of the asteroid and analyzed the stability of relative equilibrium of a spacecraft. Antoniadou and Voyatzis [21] studied the periodic orbits in the planetary system, and found that the stable periodic orbits will lead to long-term stability while the unstable orbits will lead to chaotic motion and destabilize the system.

Scheeres et al. [22, 23] presented the dynamical equation, effective potential, and the Jacobian constant around the minor celestial bodies. Yu and Baoyin [15] gave a grid search method with the hierarchical parameterization to calculate the periodic orbits and periodic orbit families around minor celestial bodies. Jiang et al. [6] derived the linearised motion equations around equilibrium points and the characteristic equation of the equilibrium points of minor celestial bodies, which is useful to calculate the stability, topological classifications, and local motions around equilibrium points. The local periodic orbits and quasi-periodic orbits can be calculated by the analytic method presented from Jiang et al. [6]. The accuracy of analytic method is lower than that of the grid search method when computing the periodic orbits around equilibrium points [22, 24]. However, the compute speed of the analytic method is faster than that of the grid search method. If one wants to calculate the global periodic orbits, one can only use the numerical method, such as the grid search method [15, 16, 18]. The periodic orbit around a minor celestial body has six characteristic multipliers, at least two of which equal 1 [7, 15, 22].

Jiang et al. [7] found four kinds of bifurcations of periodic orbits when continuing the periodic orbits, including the real saddle bifurcations, the period-doubling bifurcations, the tangent bifurcations and the Neimark-Sacker bifurcations. When continuing the periodic orbits around asteroids or comets, if the bifurcations occur, the stability of the periodic orbits may vary. Jiang and Baoyin [17] presented a conserved quantity which can restrict the number of periodic orbits on a fixed energy curved surface in the potential of a minor celestial body. They also discussed multiple bifurcations in the periodic orbit families around asteroids. Ni et al. [18] furthermore calculated several different kinds of multiple bifurcations in the periodic orbit families around asteroid 433 Eros.

Because bifurcations of stable periodic orbits may lead to unstable ones, when computing stable periodic orbits around minor celestial bodies, bifurcations are not expected to occur, especially for the design of stable periodic orbits for spacecrafts orbiting asteroids or comets, or for the study of stable periodic orbits for moonlets orbiting the primary in the binary or triple systems [3, 11, 19]. Jiang et al. [16] found a family of stable periodic orbits, which is retrograde, nearly circular, and with zero-inclinations relative to the primary's body-fixed frame. When continuing the periodic orbits, the characteristic multipliers collide at -1 and pass through each other. The period-doubling bifurcation doesn't occur during the continuation. The four characteristic multipliers except two equal to 1 are in the unit circle. During the continuation, the characteristic multipliers will collide on the unit circle, but the Neimark-Sacker bifurcation doesn't occur. Using the contents about the continuity of periodic orbits during the change of parameters from Jiang et al. [7], we know that there exists a family of stable periodic orbits around each of the stable periodic orbit. Thus here we mainly discuss the stable periodic orbits around minor celestial bodies.

This paper is organized as follows. Section 2 focuses on the gravitational potential of minor celestial bodies. Section 3 discusses the monodromy matrix and characteristic multipliers of periodic orbits as well as the stability and topological classifications of stable periodic orbits. In Section 4, we find five different kinds of stable periodic orbits in the potential of minor celestial bodies. These different kinds of stable periodic orbits are found in the potential of several minor celestial bodies, including the comet 1P/Halley, the asteroids 243 Ida, 433 Eros, 6489 Golevka, and 101955 Bennu.

## 2. Gravitational Potential

The shape and gravitational model of minor celestial bodies can be modeled by the polyhedral model [3, 25-27] or the hard/soft-sphere discrete element method [10, 28-30]. Asteroid 433 Eros is elongated, and has both concave and convex areas on surface. So we choose asteroid 433 Eros to calculate the irregular shape and effective potential. The physical and shape model of asteroid 433 Eros used here is generated by data from Gaskell [31] with the polyhedral model [25, 26]. The overall dimensions of asteroid 433 Eros are $36 \times 15 \times 13$ km [32], the estimated bulk density is 2.67 g m$^{-3}$ [32, 33], the rotational period is 5.27025547 h [32] and the moment of inertia is $17.09 \times 71.79 \times 74.49$ km$^2$ [33]. The modeling of 433 Eros employed 99846 vertices and 196608 faces [31]. Figure 1 shows the 3D Shape of asteroid 433 Eros viewed from different directions. From Figure 1, one can see that there are several craters with different size on the surface of asteroid 433 Eros. There is a large crater on the +y direction of the asteroid.

Figure 2 shows a 3-D contour plot of the effective potential $V$ for asteroid 433 Eros. The values for $V$ are plotted in the xy, yz, and zx planes, respectively. The structures of effective potential in the xy, yz, and zx planes are quite different. More detailed contents to calculate the effective potential can be found in Jiang and Baoyin [17]. The positions of equilibrium points of asteroid 433 Eros is near the xy plane, but not in the xy plane, which implies that they are out-of-plane equilibrium points. More details about the equilibrium points of asteroids can be found in Chanut et al. [4], Jiang et al. [6], and Wang et al. [27].

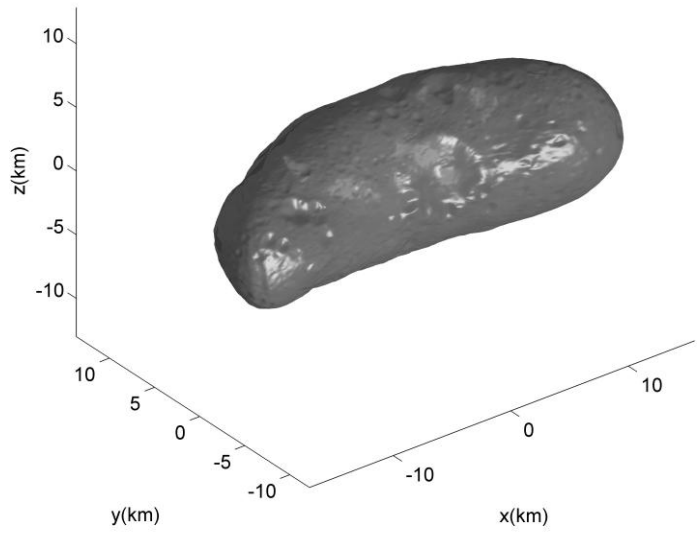

(a)

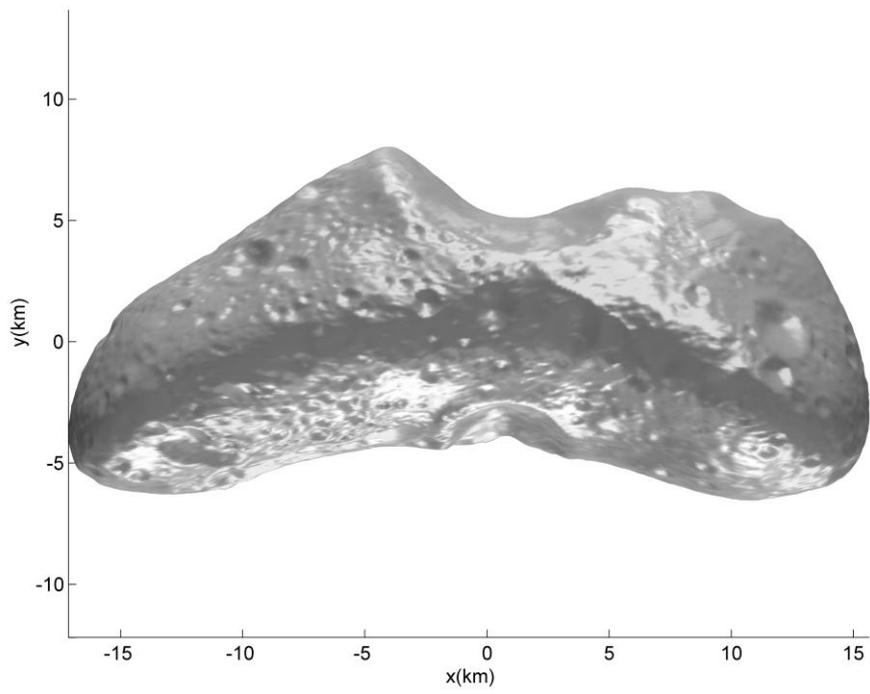

(b)

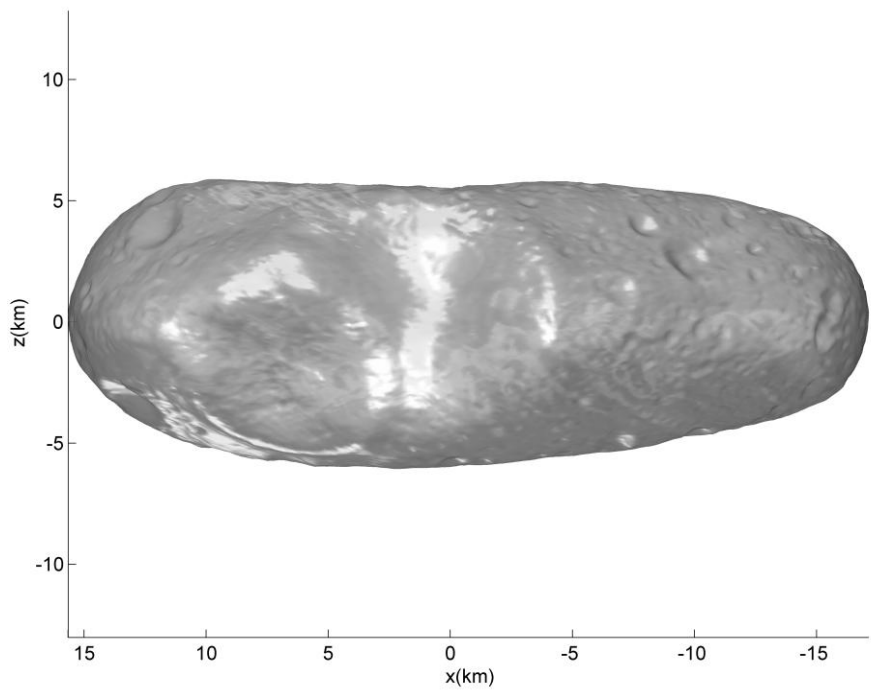

(c)

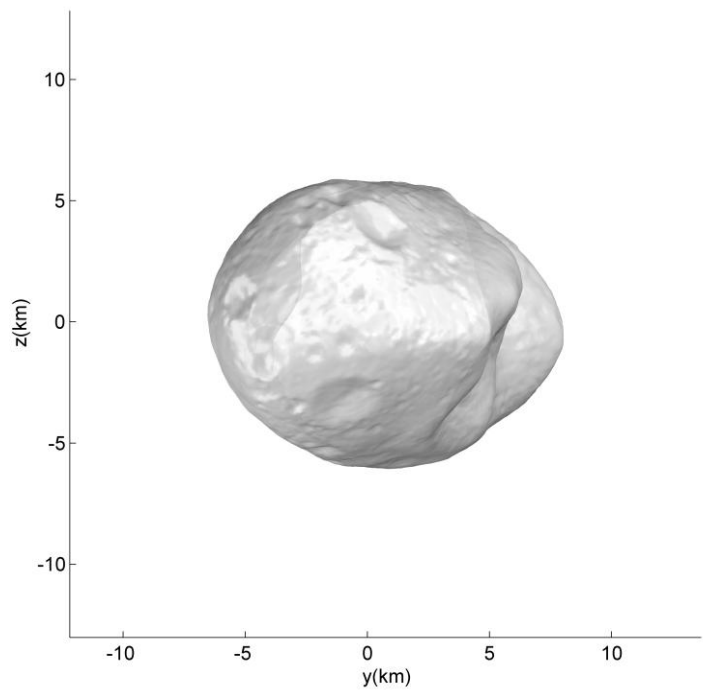

(d)

Figure 1. 3D Shape of asteroid 433 Eros.

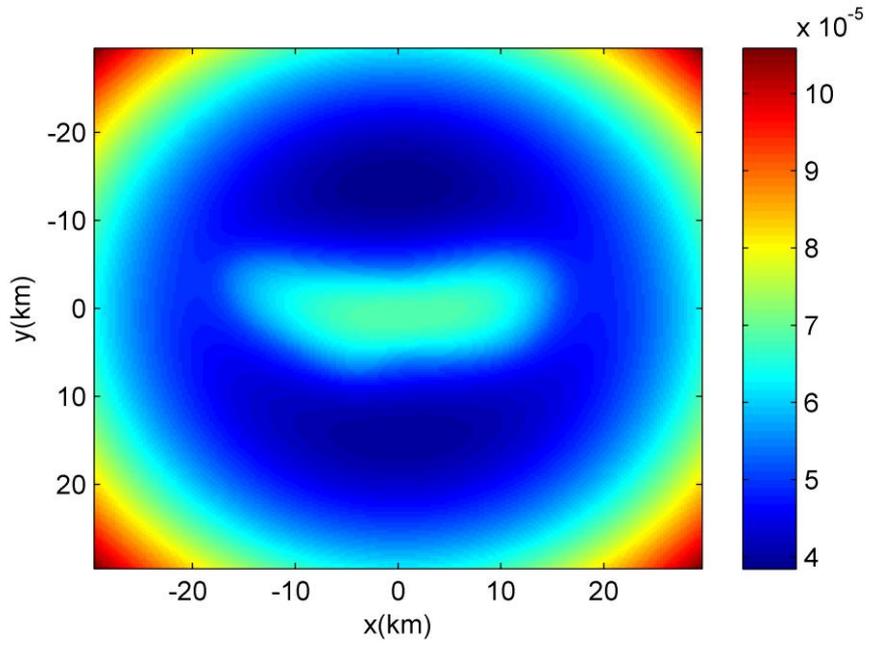

(a)

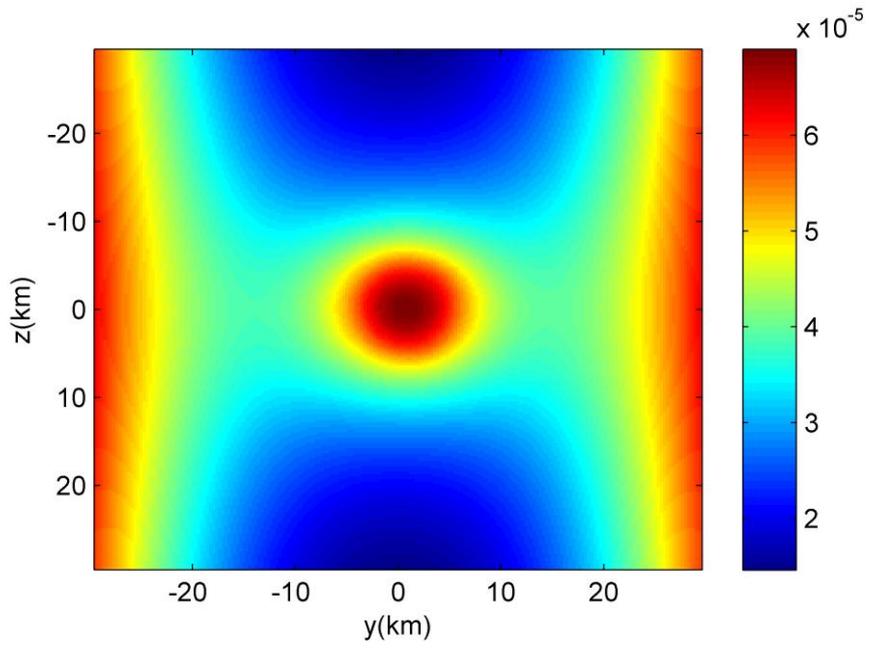

(b)

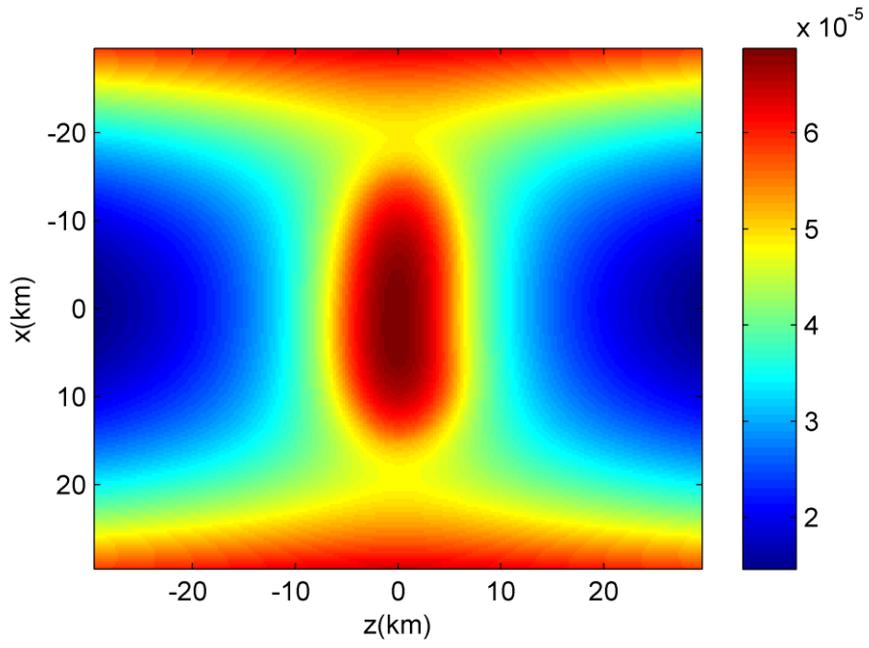

(c)

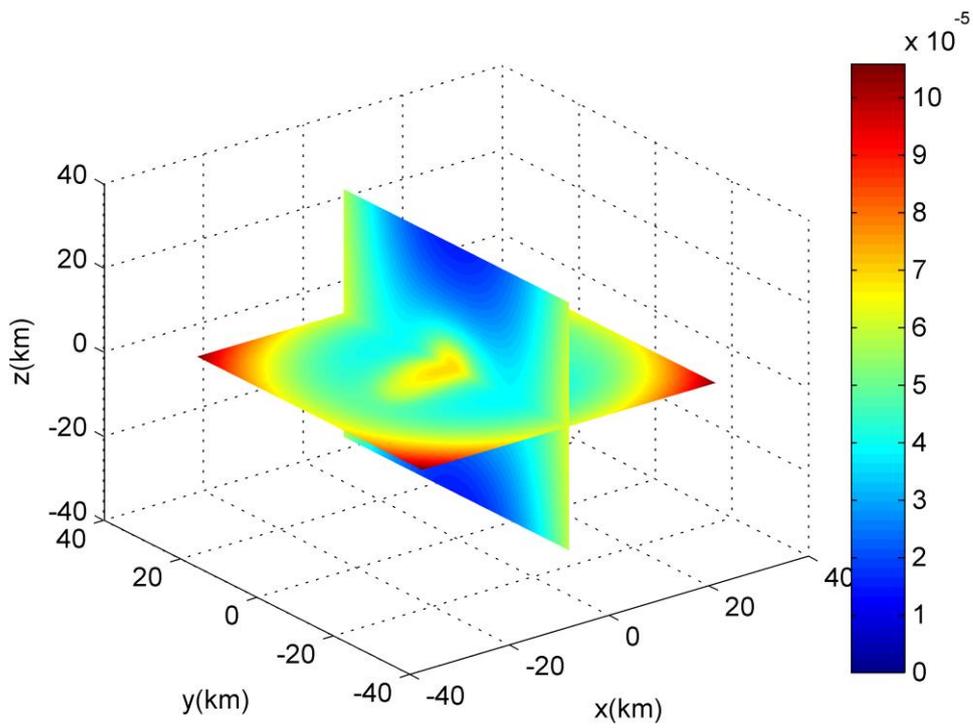

(d)

Figure 2. A 3-D contour plot of the effective potential for 433 Eros (unit: km$^2$ s$^{-2}$).

## 3. Stable Periodic Orbits for Spacecrafts around Asteroids and Comets

### 3.1 Monodromy Matrix and Characteristic Multipliers

The dynamical equation of the spacecraft relative to the body-fixed frame of the minor celestial body can be expressed in the symplectic form [7]

$$\dot{\mathbf{z}} = \mathbf{F}(\mathbf{z}) = \mathbf{J}\nabla H(\mathbf{z}), \tag{1}$$

where $\mathbf{z} = [\mathbf{p}\ \mathbf{q}]^T$, $\mathbf{p} = (\dot{\mathbf{r}} + \boldsymbol{\omega} \times \mathbf{r})$ is the generalised momentum, $\mathbf{q} = \mathbf{r}$ is the generalised coordinate, $\mathbf{q}$ represents the position vector of the spacecraft relative to the body's body-fixed frame, $\mathbf{J} = \begin{pmatrix} \mathbf{0} & -\mathbf{I} \\ \mathbf{I} & \mathbf{0} \end{pmatrix}$, $\mathbf{I}$ is a $3 \times 3$ unit matrix, $\mathbf{0}$ is a $3 \times 3$ zero matrix, $H = -\frac{\mathbf{p} \cdot \mathbf{p}}{2} + U(\mathbf{q}) + \mathbf{p} \cdot \dot{\mathbf{q}}$ is the Hamilton functions, $U(\mathbf{q})$ respresents the body's gravitational potential, and $\nabla H(\mathbf{z}) = \left(\frac{\partial H}{\partial \mathbf{p}}\ \frac{\partial H}{\partial \mathbf{q}}\right)^T$ is the gradient of $H(\mathbf{z})$. $U(\mathbf{q})$ can be computed by the polyhedral model [26] with the shape data and physical parameters of the minor celestial body. The body-fixed frame is defined by the inertia axis, i.e. x, y, and z axes are the minimum, medium, and maximum inertia axis, respectively.

Denote $S_p(T)$ as the set of periodic orbits with the period $T$. For each periodic orbit $p \in S_p(T)$, it has a $6 \times 6$ matrix $\nabla \mathbf{F} := \frac{\partial \mathbf{F}(\mathbf{z})}{\partial \mathbf{z}}$, then the state transition matrix [7, 15, 21, 34] for the periodic orbit can be written as

$$\Phi(t) = \int_0^t \frac{\partial \mathbf{f}}{\partial \mathbf{z}}(p(\tau))d\tau, \tag{2}$$

the periodic orbit's monodromy matrix is then

$$M = \Phi(T). \tag{3}$$

Characteristic multipliers of the periodic orbit are the eigenvalues of the monodromy matrix. Each periodic orbit has six characteristic multipliers, and all the characteristic multipliers take the form of $e^{\pm\sigma\pm i\tau}$ $(\sigma,\tau\in\mathrm{R};\sigma>0,\tau\in(0,\pi))$, $\mathrm{sgn}(\alpha)e^{\pm\alpha}$ $(\alpha\in\mathrm{R},|\alpha|\in(0,1))$, $e^{\pm i\beta}$ $(\beta\in(0,\pi))$, $-1$, and 1, where

$$\mathrm{sgn}(\alpha)=\begin{cases}1 & (\text{if } \alpha>0)\\ -1 & (\text{if } \alpha<0)\end{cases}.$$

**3.2 Stability of Periodic Orbits**

Distribution of six characteristic multipliers of the periodic orbit determines the topological classifications of periodic orbits [35]. The topological classifications of stable periodic orbits [17] have 7 different cases, which is listed in Table 1.

Table 1. The topological classifications of stable periodic orbits

| Topological cases | Characteristic multipliers |
|---|---|
| Case N2 (Non-degenerate cases) | $e^{\pm i\beta_j}$ $(\beta_j\in(0,\pi); j=1,2\|\beta_1\neq\beta_2)$, $\gamma_j$ $(\gamma_j=1; j=1,2)$ |
| Case DP1 (Degenerate periodic cases) | $e^{\pm i\beta_j}$ $(\beta_j\in(0,\pi); j=1)$, $\gamma_j$ $(\gamma_j=1; j=1,2,3,4)$ |
| Case DP2 (Degenerate periodic cases) | $\gamma_j$ $(\gamma_j=1; j=1,2,3,4,5,6)$ |
| Case K1 (Krein collision cases) | $e^{\pm i\beta_j}$ $(\beta_j\in(0,\pi); j=1,2\|\beta_1=\beta_2)$, $\gamma_j$ $(\gamma_j=1; j=1,2)$ |
| Case PD1 (Period-doubling cases) | $\gamma_j$ $(\gamma_j=1; j=1,2,3,4)$, $\gamma_j$ $(\gamma_j=-1; j=1,2)$ |
| Case PD2 (Period-doubling cases) | $\gamma_j$ $(\gamma_j=1; j=1,2)$, $\gamma_j$ $(\gamma_j=-1; j=1,2,3,4)$ |
| Case PD3 (Period-doubling cases) | $\gamma_j$ $(\gamma_j=1; j=1,2)$, $\gamma_j$ $(\gamma_j=-1; j=1,2), e^{\pm i\beta_j}$ $(\beta_j\in(0,\pi); j=1)$ |

More detailed contents of the topological classifications of periodic orbits can be seen in Jiang and Baoyin [17]. The topological case of the periodic orbits in the periodic orbit family may vary from the stable cases to the unstable cases. There are three bifurcations [7][17, 18] related to the variety of the stability of the periodic orbits, i.e. Neimark-Sacker bifurcations, tangent bifurcations, as well as period-doubling bifurcations.

**4. Different Kinds of Stable Periodic Orbits**

In this section, we present five different kinds of stable periodic orbits around minor celestial bodies. These periodic orbits are calculated using the grid search method developed by Yu and Baoyin [15]. Before the periodic orbits are presented, we first give the length unit and time unit used here in Table 2. For instance, the first line in Table 2 means that the length unit for motion around comet 1P/Halley is defined to be 15.140166km, and the time unit is defined to be 52.8 h.

Table 2. The length unit and time unit used in this paper for minor bodies

| Minor bodies | length unit | time unit |
|---|---|---|
| comet 1P/Halley | 15.140166km | 52.8h |
| 243 Ida | 57.7917km | 4.63h |
| 433 Eros | 34.4km | 5.2656h |
| 101955 Bennu | 566.4413m | 4.288h |
| 6489 Golevka | 685.15093m | 6.026h |

**4.1 Generated from the stable equilibrium points**

If the minor celestial bodies have stable equilibrium points outside the bodies, there exist three families of stable periodic orbits around each of the stable equilibrium points. There are several asteroids and comets having stable equilibrium points outside, including asteroid 4 Vesta [27], 2867 Steins [27], 6489 Golevka [6], 52760 [27], and comets 1P/Halley [9] as well as 9P/Tempel1 [27]. Jiang et al. [6] presented the analytic method to calculate the local periodic orbits around stable equilibrium points; however, the local periodic orbits calculated from the analytic method are not accurate enough. Using the grid search method developed by Yu and Baoyin [15], the stable periodic orbits around the stable equilibrium points can be calculated numerically. The grid search method has a high accuracy. The analytic method from Jiang et al. [6] can be used to give an initial estimate for the grid search method developed by Yu and Baoyin [15]. Jiang [24] applied the grid search method to the computation of the local periodic orbits around equilibrium points of asteroid 216 Kleopatra. Here we discuss the stable periodic orbits generated from the stable equilibrium points of minor celestial bodies. We choose comet 1P/Halley to calculate the stable periodic orbits around the stable equilibrium points. We use the grid search method to search the periodic orbits and Eq. (3) to calculate the distribution of six characteristic multipliers of the periodic orbits. There are totally four equilibrium points outside the body of the comet 1P/Halley, two of them are stable, i.e. E2 and E4 [7, 27].

Figure 3 shows a periodic orbit continued from the stable equilibrium point E2 (The positions and serial numbers of equilibrium points in this paper can be seen in Wang et al. [7]) of the comet 1P/Halley. The period of this periodic orbit is 53.3971h.

The rotation period of the comet 1P/Halley is 52.8h. Thus the ratio of the period of the periodic orbit relative to the period of the comet is 1.01130878575. From Figure 3(b), one can see that the periodic orbit is stable.

Figure 4 shows a periodic orbit continued from the stable equilibrium point E4 of the comet 1P/Halley. The period of this periodic orbit is 53.3851h. The ratio of the period of the periodic orbit and the period of the comet is 1.011080562545. From Figure 4(b), one can see that the periodic orbit is stable. Table 3 presents the initial positions and the initial velocities of these two periodic orbits presented in Figure 3 and 4, the values are expressed in the body-fixed frame of 1P/ Halley. In Table 3, periodic orbit 1 corresponds to the orbit presented in Figure 3 while periodic orbit 2 corresponds to the orbit presented in Figure 4.

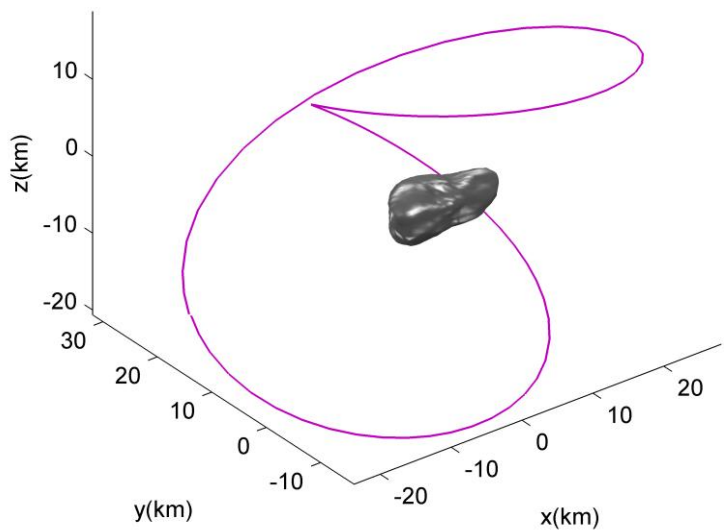

(a)

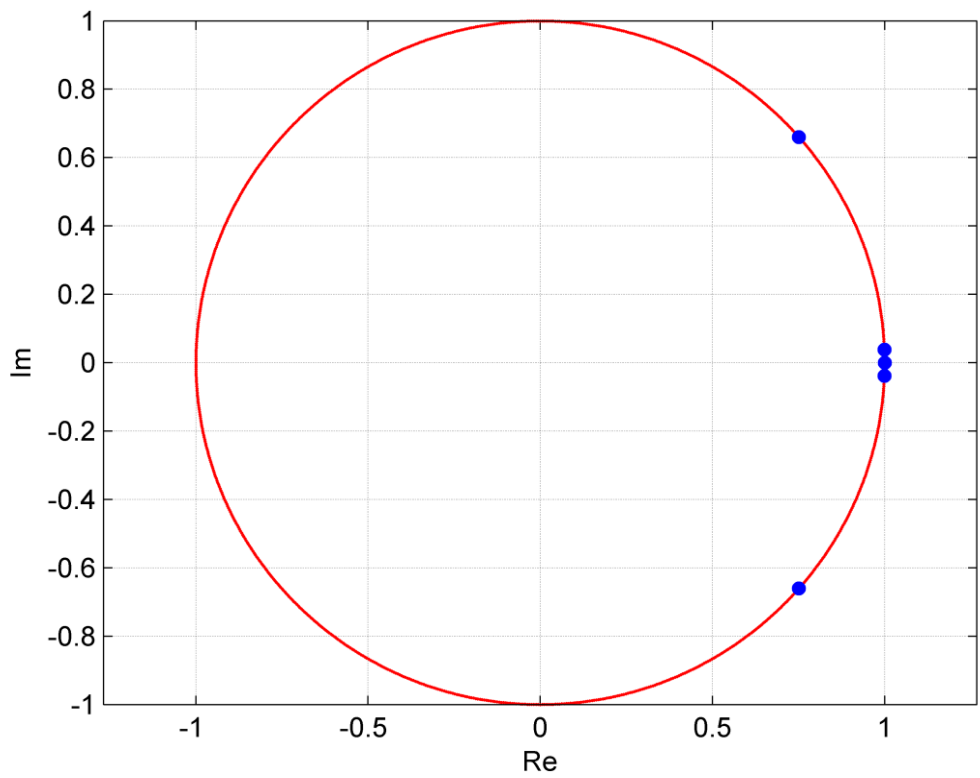

(b)

Figure 3. A periodic orbit continued from the stable equilibrium point E2 of the comet 1P/Halley, the period is 53.3971h, and the ratio of the period of the periodic orbit and the period of the comet is 1.01130878575. (a) The 3D plot of the periodic orbit relative to the body-fixed frame of 1P/Halley; (b) The distribution of six characteristic multipliers of the periodic orbit.

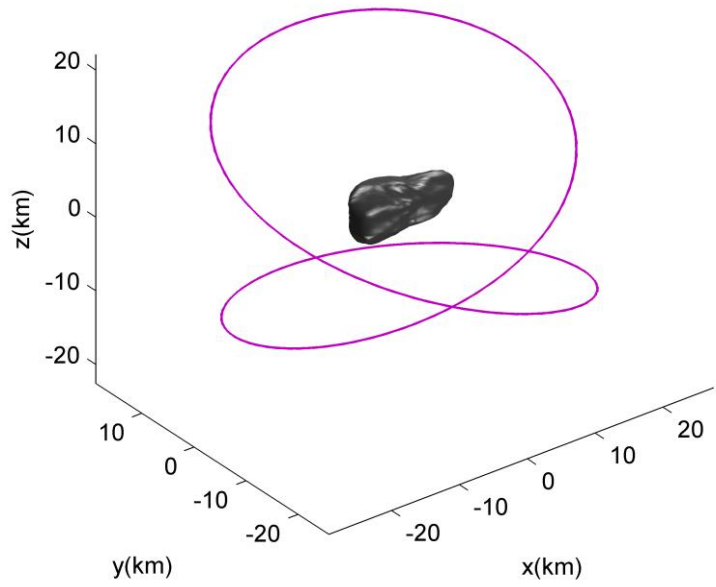

(a)

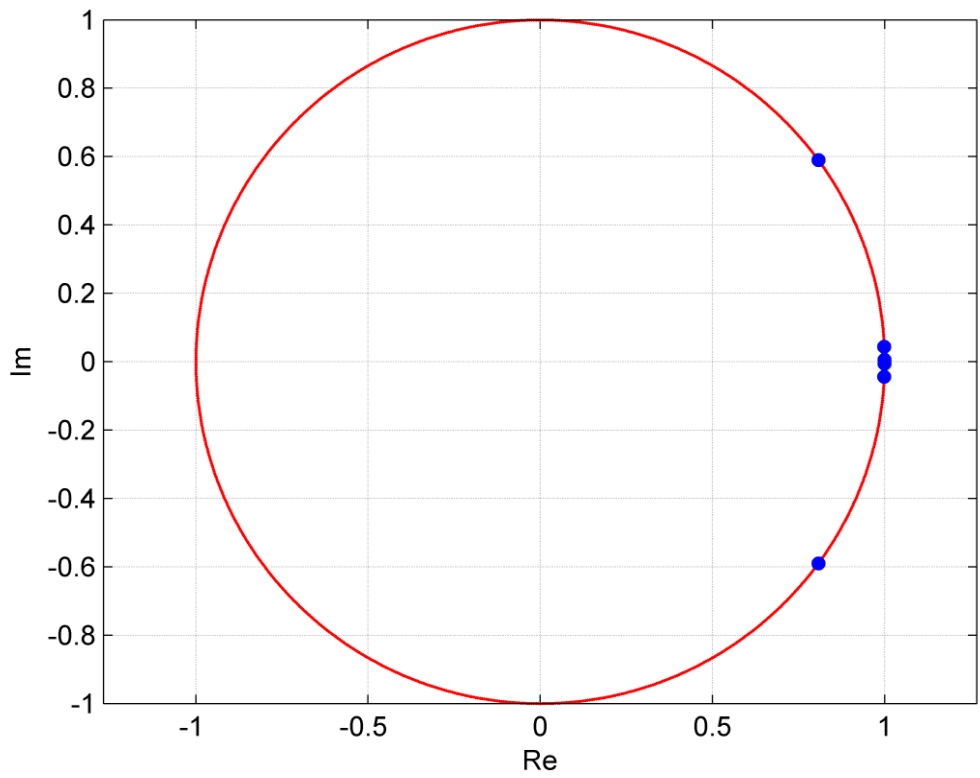

(b)

Figure 4. A periodic orbit continued from the stable equilibrium point E4 of the comet 1P/Halley, the period is 53.385053702375991h, and the ratio of the period of the periodic orbit and the period of the comet is 1.011080562545. (a) The 3D plot of the periodic orbit relative to the body-fixed frame of 1P/Halley; (b) The distribution of six characteristic multipliers of the periodic orbit.

Table 3. The initial positions and the initial velocities of periodic orbits in the body-fixed frame of 1P/ Halley

| Periodic Orbits | Positions | Velocities | Periods |
|---|---|---|---|
| 1 | -1.43509739861<br>1.10624686366<br>-0.945745920199 | 6.58815083642<br>15.2096319563<br>4.79820012348 | 1.01130878575 |
| 2 | 1.44026951894<br>-0.116377265356<br>-0.913778421911 | 1.26316060312<br>-17.8320567684<br>3.71316717169 | 1.011080562545 |

## 4.2 Continued from the unstable periodic orbits around the unstable equilibrium points

The distribution of eigenvalues determines the topological cases of the equilibrium points around a uniformly rotating body. Jiang et al. [6] classified several different topological cases of the equilibrium points. From Jiang et al. [6] and Wang et al. [27], we know that for most minor celestial bodies, if the external equilibrium points are unstable, they belong to Case 2 or Case 5; where Case 2 has the eigenvalues $\pm \alpha_j \left( \alpha_j \in \mathbf{R}^+; j=1 \right)$ and $\pm i\beta_j \left( \beta_j \in \mathbf{R}^+; j=1,2 \right)$ while Case 5 has the eigenvalues $\pm \sigma \pm i\tau \left( \sigma, \tau \in \mathbf{R}^+ \right)$ and $\pm i\beta_j \left( \beta_j \in \mathbf{R}^+; j=1 \right)$. In the vicinity of the equilibrium points which belong to Case 2, there exist two families of unstable periodic orbits. In the vicinity of the equilibrium points which belong to Case 5, there exists one family of unstable periodic orbits. When continuing these unstable periodic orbits, the amplitude of the periodic orbits increase gradually, and the periodic orbits become stable.

Here we choose the comet 1P/Halley and the primary of the binary asteroid 243 Ida to calculate the stable periodic orbits continued from the unstable periodic orbits around the unstable equilibrium points. Comet 1P/Halley has two unstable equilibrium points E1 and E3. The primary of the binary asteroid 243 Ida has four equilibrium points outside the body. All of them are unstable.

Figure 5 and 6 shows two periodic orbits continued from the unstable equilibrium point E1 and E3 of the comet 1P/Halley, respectively. The period of the periodic orbit presented in Figure 5 is 53.454h, the ratio of the period of the periodic orbit and the period of the comet is 1.01238521471. The period of the periodic orbit presented in Figure 6 is 54.0119h, the ratio of the period of the periodic orbit relative to the period of the comet is 1.02295314063. From Figure 5(b) and Figure 6(b), one can see that these two periodic orbits are stable.

Figure 7 and 8 shows two periodic orbits continued from the unstable equilibrium point E2 and E4 of the primary of the binary asteroid 243 Ida, respectively. The period of the periodic orbit presented in Figure 7 is 5.2447h, the ratio of the period of the periodic orbit and the period of the comet is 1.13276985098. The periodic orbit presented in Figure 7 is continued from the unstable equilibrium point E2, it connects periodic orbit generated from equilibrium point E1. The period of the periodic orbit presented in Figure 8 is 5.2678h, the ratio of the period of the periodic orbit and the period of the comet is 1.13775669387. From Figure 7(b) and Figure 8(b), one can see that these two periodic orbits are stable.

Table 4 presents the initial positions and the initial velocities of these four periodic orbits presented in Figure 5-8, the values are expressed in the body-fixed frame of the minor bodies. In Table 4, periodic orbits 1 and 2 correspond to the orbits presented in Figure 5 and 6, respectively; periodic orbits 3 and 4 correspond to the orbits presented in Figure 7 and 8, respectively.

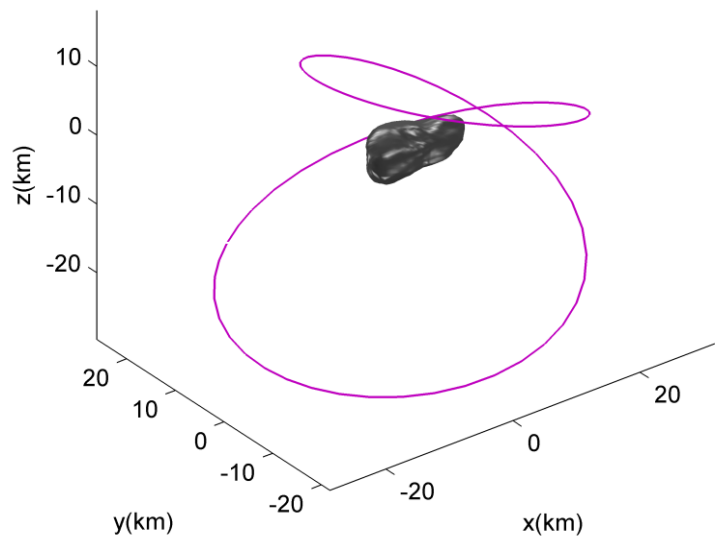

(a)

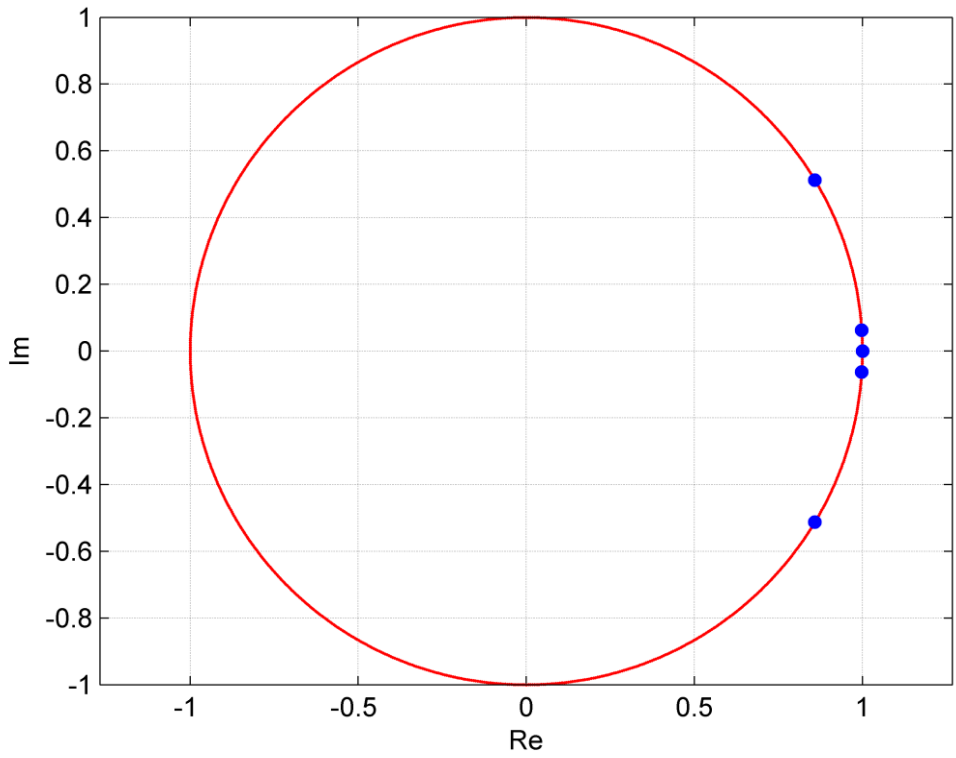

(b)

Figure 5. A periodic orbit continued from the stable equilibrium point E1 of the comet 1P/Halley, the period is 53.454h, and the ratio of the period of the periodic orbit and the period of the comet is 1.01238521471. (a) The 3D plot of the periodic orbit relative to the body-fixed frame of 1P/Halley; (b) The distribution of six characteristic multipliers of the periodic orbit.

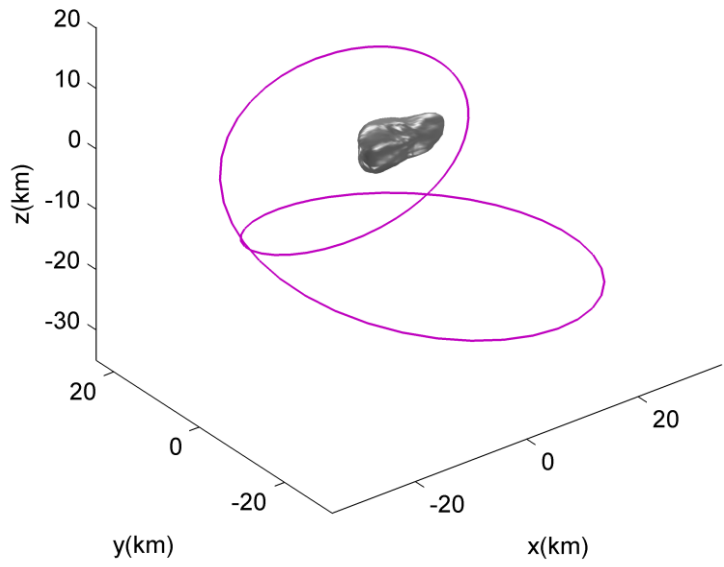

(a)

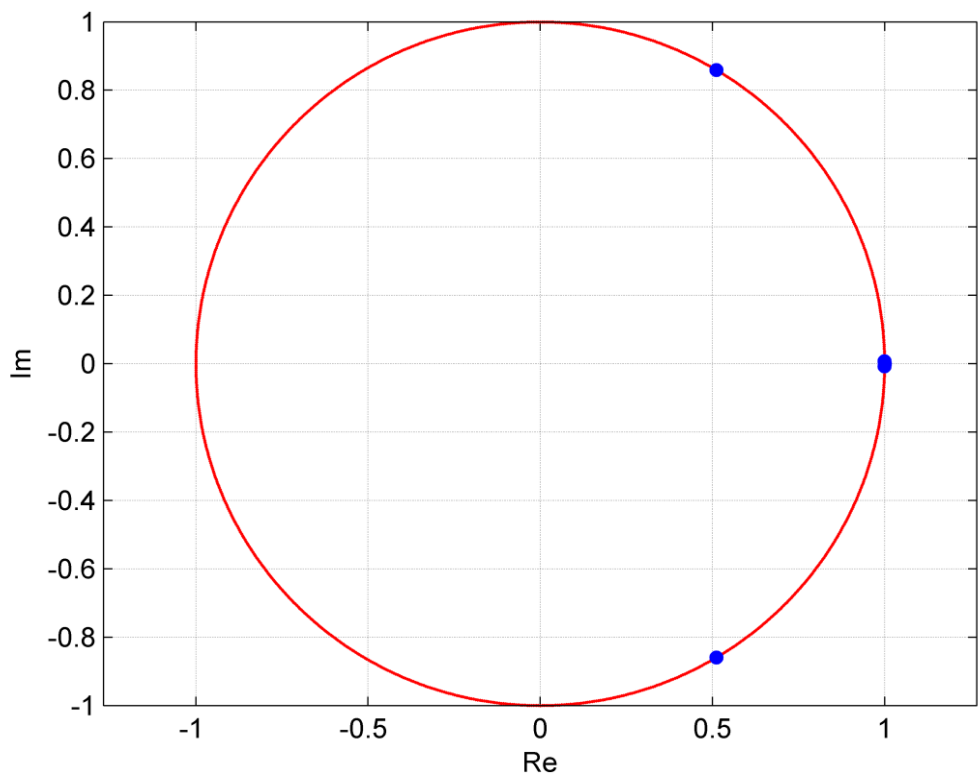

(b)

Figure 6. A periodic orbit continued from the stable equilibrium point E3 of the comet 1P/Halley, the period is 54.011925825h, and the ratio of the period of the periodic orbit and the period of the comet is 1.02295314063. (a) The 3D plot of the periodic orbit relative to the body-fixed frame of 1P/Halley; (b) The distribution of six characteristic multipliers of the periodic orbit, there are four multipliers equal 1.

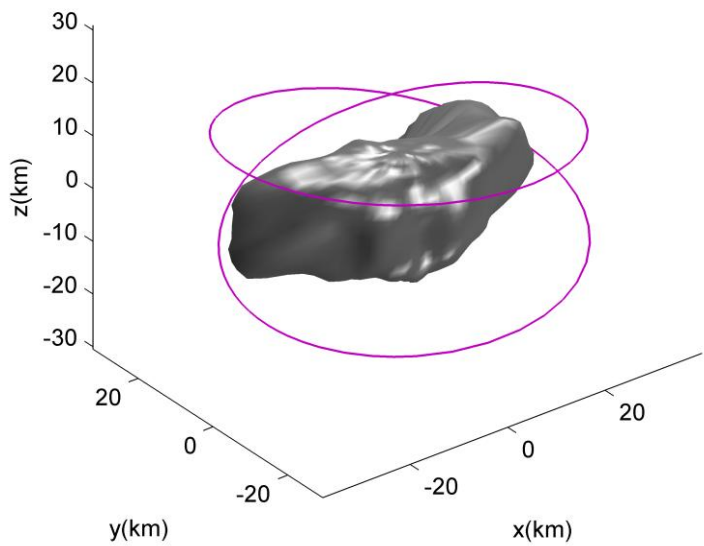

(a)

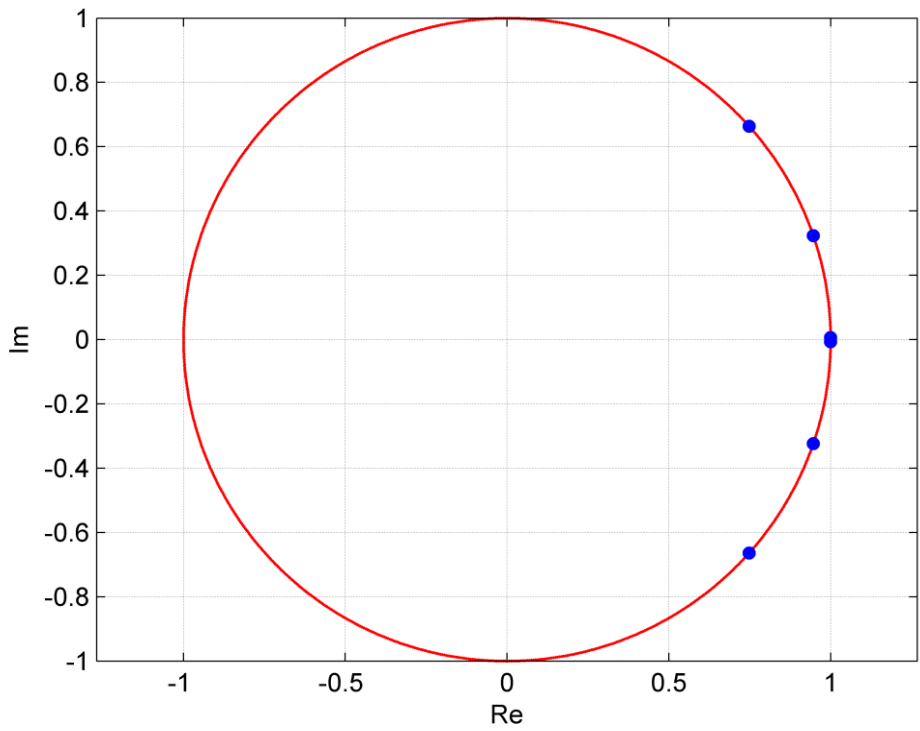

(b)

Figure 7. A periodic orbit continued from the stable equilibrium point E2 of the asteroid 243 Ida, the period is 5.244724410037399h, and the ratio of the period of the periodic orbit and the period of the asteroid is 1.13276985098. (a) The 3D plot of the periodic orbit relative to the body-fixed frame of 243 Ida; (b) The distribution of six characteristic multipliers of the periodic orbit.

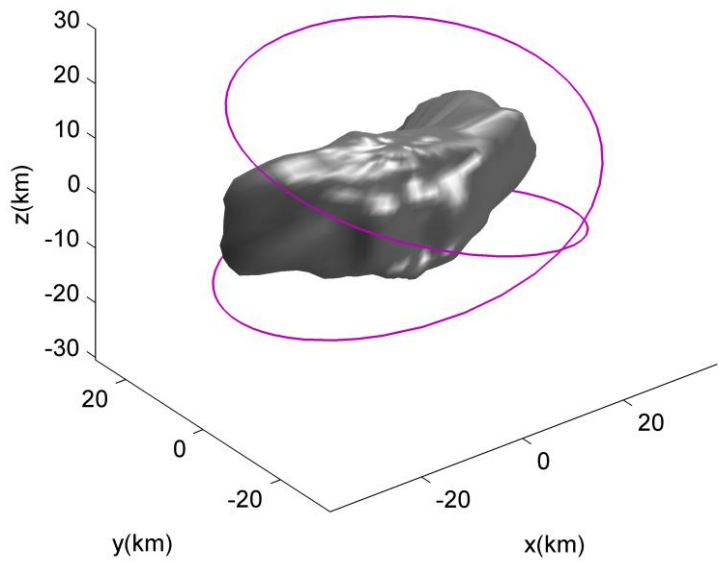

(a)

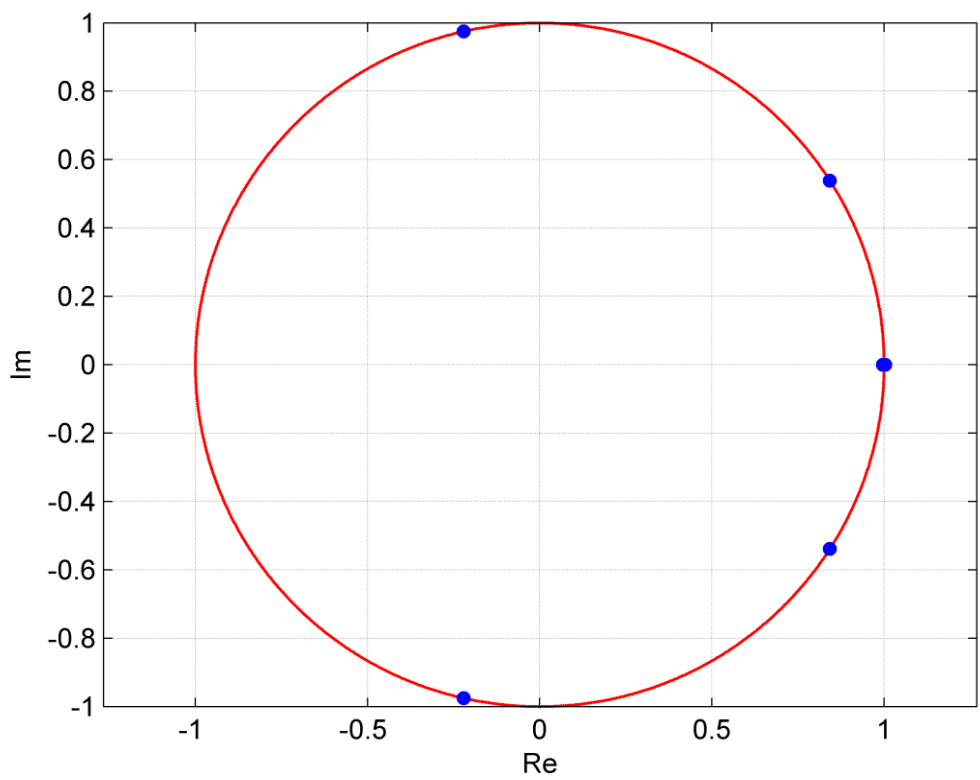

(b)

Figure 8. A periodic orbit continued from the stable equilibrium point E4 of the asteroid 243 Ida, the period is 5.2678134926181h, and the ratio of the period of the periodic orbit and the period of the asteroid is 1.13775669387. (a) The 3D plot of the periodic orbit relative to the body-fixed frame of 243 Ida; (b) The distribution of six characteristic multipliers of the periodic orbit.

Table 4. The initial positions and the initial velocities of periodic orbits in the body-fixed frame of 1P/Halley and 243 Ida

| Periodic Orbits | Positions | Velocities | Periods |
|---|---|---|---|
| 1 | -0.724053725841<br>1.48667560655<br>-1.33198379992 | 14.3328568184<br>8.78463438036<br>3.48528278685 | 1.01238521471 |
| 2 | 0.0508067344449<br>1.58710718406<br>-1.43117795230 | 16.7180042361<br>1.28433682355<br>-2.60834460269 | 1.02295314063 |
| 3 | 0.0626044457277<br>0.532671731883<br>-0.0571937606260 | 5.70149686669<br>-0.332300886588<br>1.49480516557 | 1.13276985098 |
| 4 | 0.365717153565<br>-0.376264464591<br>-0.100156740647 | -3.96356504016<br>-3.96490101038<br>1.66824049654 | 1.13775669387 |

**4.3 Retrograde and nearly circular periodic orbits with zero inclination**

Jiang et al. [16] found a family of stable periodic orbits around the large-size-ratio triple asteroid 216 Kleopatra, which is nearly circular and retrograde with zero inclination. These periodic orbits are nearly circular and retrograde with zero inclination. The primary of 216 Kleopatra has seven equilibrium points, four of them are outside the body, and these four equilibrium points are all unstable. We choose some other minor celestial bodies with different structure of gravitational fields to calculate if similar orbits also exist. Asteroid 433 Eros, 101955 Bennu and 6489 Golevka are taken for examples to calculate the periodic orbits. 433 Eros has totally five equilibrium points. Four of them are outside the body, and all the external equilibrium points are unstable. 6489 Golevka also has totally five equilibrium points. Four of them are outside the body, however, two of the outside equilibrium points are unstable and other two outside equilibrium points are stable. 101955 Bennu has totally nine equilibrium points. These three asteroids have different structures of gravitational fields.

Figure 9 shows a family of periodic orbits around asteroid 433 Eros, the periodic orbits are nearly circular and retrograde with zero inclination. The initial positions and velocities of these five periodic orbits in the periodic orbit family in the gravitational potential of 433 Eros are presented in Table 5. The vectors are expressed in the asteroid's body-fixed frame. There are totally five periodic orbits plotted in Figure 9. The period of the smallest periodic orbit is 7.0827h with the ratio 1.345096. The period of the biggest periodic orbit is 7.6912h with the ratio 1.460648. Here the ratio means the ratio of the period of the periodic orbit relative to the period of the minor celestial body. From the inner to the outer, the periods of the periodic orbits become larger. This result is different from the period in the Kepler motion expressed in the inertial coordinate system. From Figure 9(b), one can see that the orbits are not in the plane; they are curves in the normal direction.

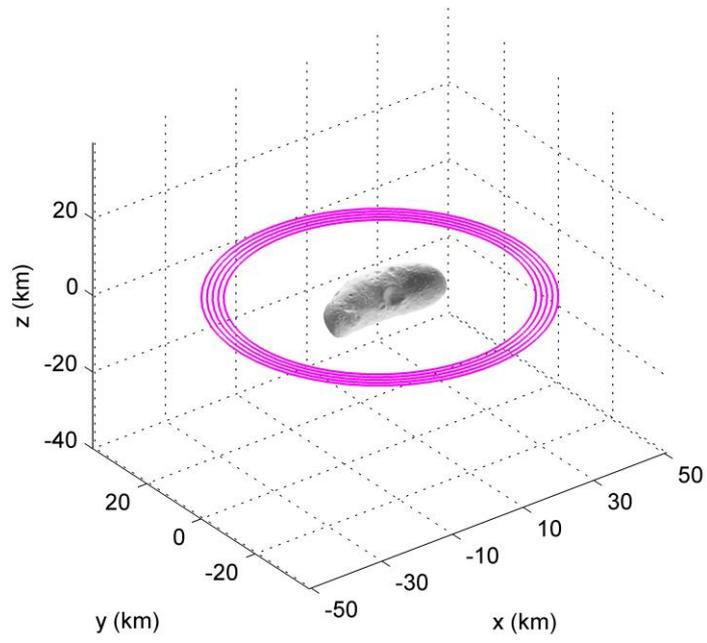

(a)

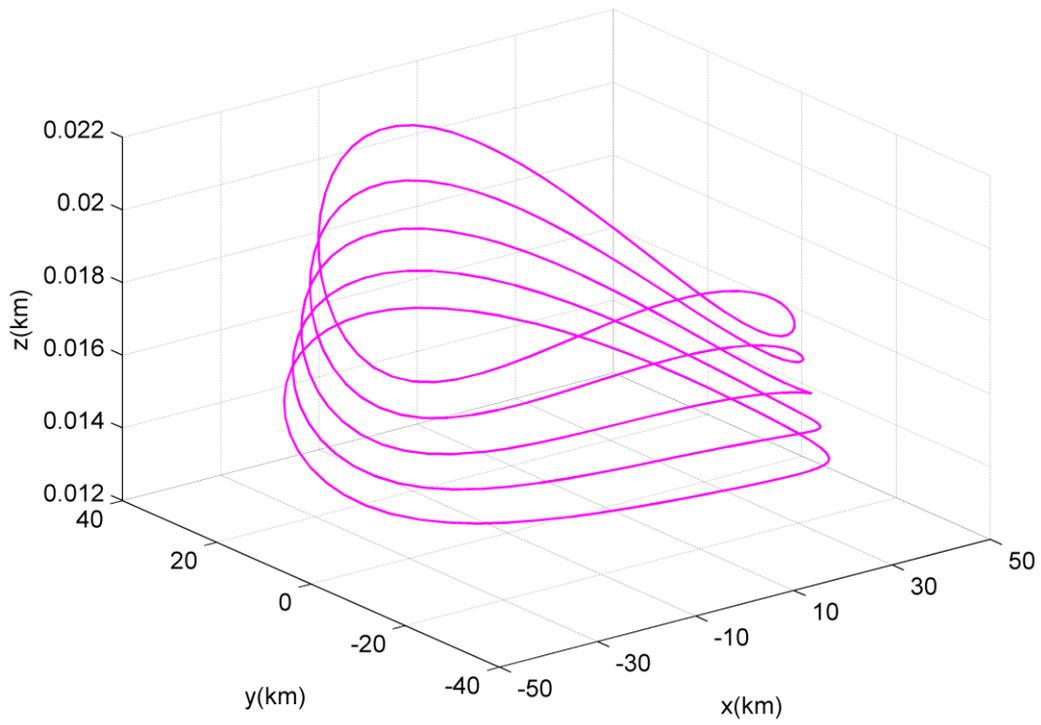

(b)

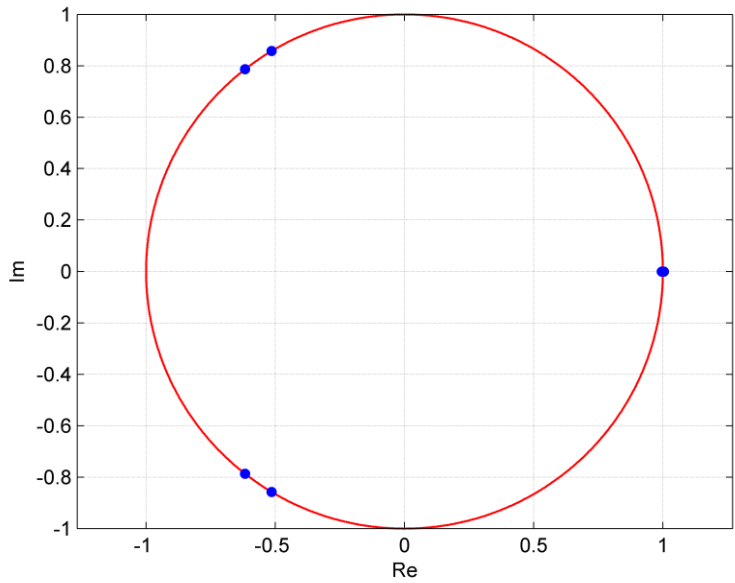

**(c)**

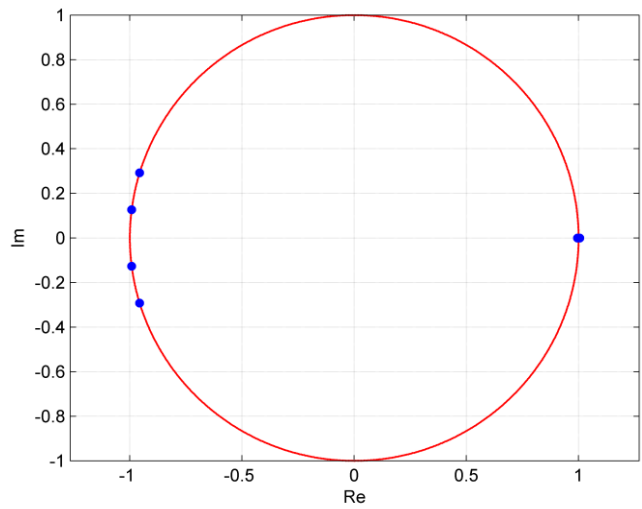

**(d)**

Figure 9. A family of nearly circular periodic orbits around asteroid 433 Eros which are retrograde with zero inclination. (a) The 3D plot of the periodic orbits relative to the body-fixed frame of 433 Eros, five periodic orbits are plotted; from the inner to the outer, the periods of the periodic orbits become bigger; (b) The 3D plot of the periodic orbits (without the asteroid) relative to the body-fixed frame of 433 Eros; (c) Distribution of six characteristic multipliers of the periodic orbit (the smallest periodic

orbit among the five periodic orbits plotted), the period is 7.0827h, the ratio of the period of the periodic orbit and the period of the asteroid is 1.345096; (d) Distribution of six characteristic multipliers of the periodic orbit (the biggest periodic orbit among the five periodic orbits plotted), the period is 7.6912h, the ratio of the period of the periodic orbit and the period of the asteroid is 1.460648.

Table 5. The initial positions and the initial velocities of periodic orbits in the body-fixed frame of 433 Eros (index 1 to 5 means the orbits from inner to outer)

| Periodic Orbits | Positions | Velocities | Periods |
|---|---|---|---|
| 1 | 0.743990716373<br>-0.894842961577<br>0.000419947168 | -4.19745020860<br>-3.44719333572<br>0.000221072321 | 1.34509555159 |
| 2 | 0.683634268907<br>-0.895480402681<br>0.000448414280 | -4.12698424359<br>-3.10294685101<br>0.000263961275 | 1.36878554932 |
| 3 | 0.614417150401<br>-0.899312199598<br>0.000480602904 | -4.06181001461<br>-2.72236601436<br>0.000314113036 | 1.39553123024 |
| 4 | 0.595998913791<br>-0.866964082507<br>0.000511515830 | -3.83587690126<br>-2.57348128953<br>0.000387281366 | 1.42591950706 |
| 5 | 0.547342936704<br>-0.854052941398<br>0.000547689132 | -3.68645740335<br>-2.28916792332<br>0.000473375432 | 1.46064834302 |

Figure 10 shows another periodic orbit family around asteroid 433 Eros. The periodic orbits are also nearly circular and retrograde with zero inclination. The initial positions and velocities of these five periodic orbits in the gravitational potential of 433 Eros are presented in Table 6. The vectors are expressed in the asteroid's body-fixed frame. There are totally seven periodic orbits plotted in Figure 9. The periodic orbits are stable. Figure 10(c) and 10(d) present the distribution of six characteristic multipliers of the smallest and biggest periodic orbits among the seven periodic orbits plotted. From Figure 10(c) and 10(d), one can see that the characteristic multipliers on the unit circle move during the continuation.

The period of the smallest periodic orbit is 8.656684h with the ratio 1.64400724599. The period of the biggest periodic orbit is 6.684357h with the ratio 1.26943881082. From the inner to the outer, the periods of the periodic orbits become smaller. This result is different from the above results presented in Figure 9 and Table 5. Thus the periodic orbit family presented in Figure 10 is different from the periodic orbit family presented in Figure 9. We can conclude that there are at least two periodic orbit families in the potential of one minor celestial body, which are nearly circular, retrograde and stable with zero inclination.

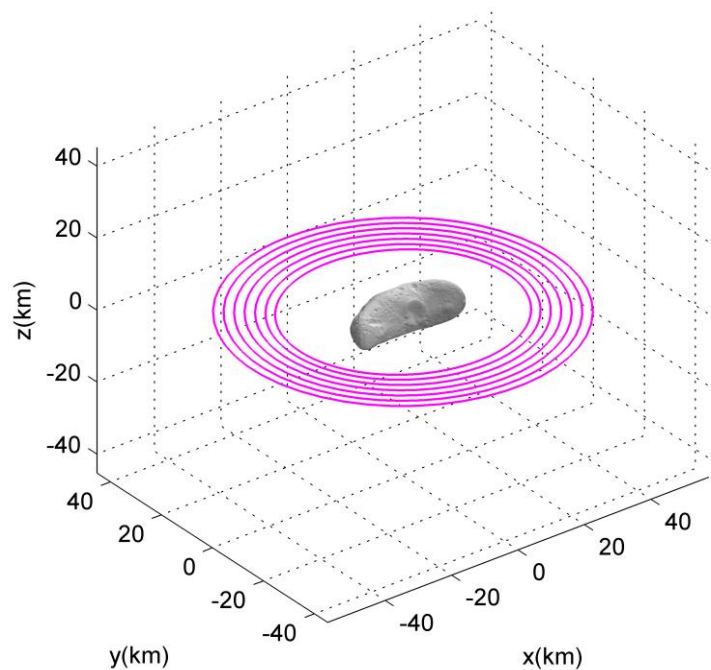

**(a)**

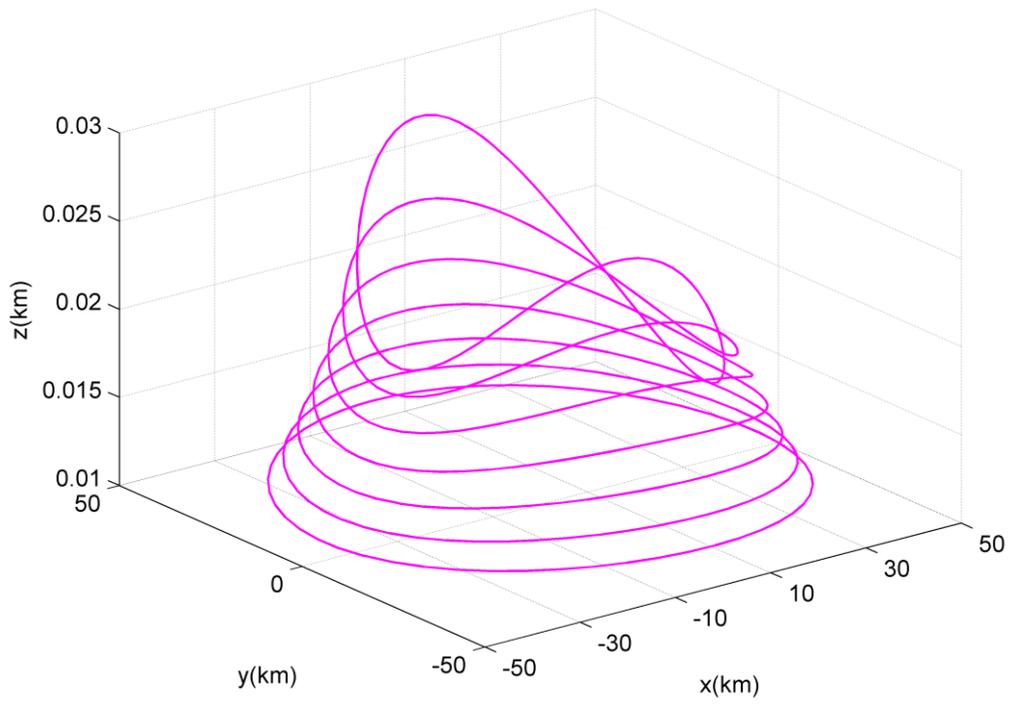

**(b)**

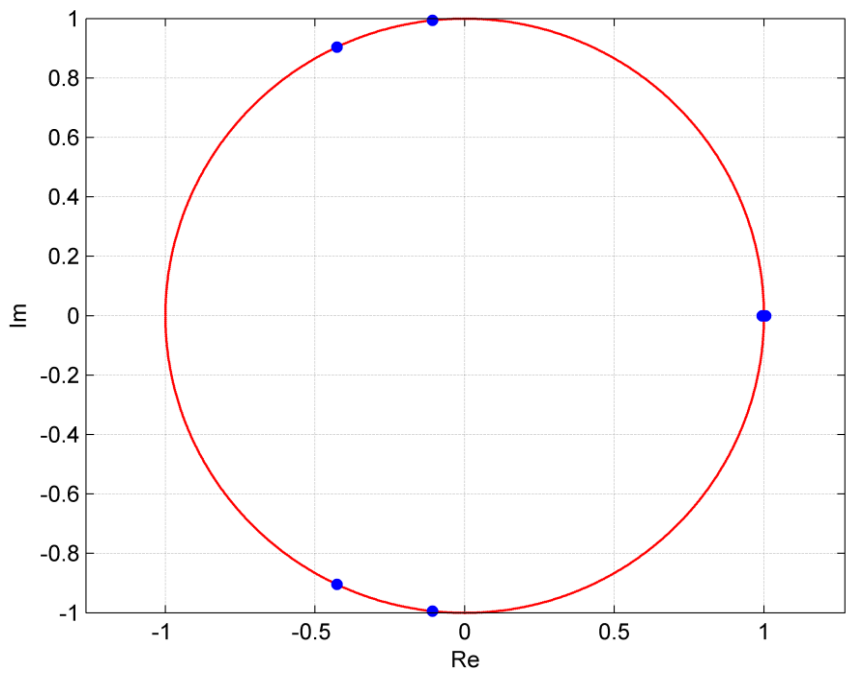

**(c)**

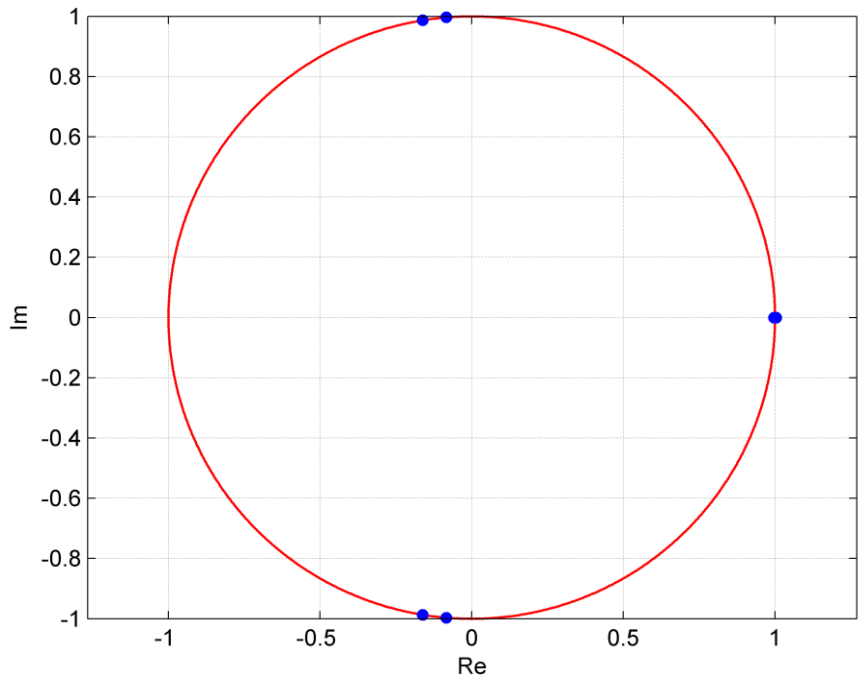

**(d)**

Figure 10. A family of nearly circular periodic orbits around asteroid 433 Eros which are retrograde with zero inclination. (a) The 3D plot of the periodic orbits relative to the body-fixed frame of 433 Eros, seven periodic orbits are plotted; from the inner to the outer, the periods of the periodic orbits become smaller; (b) The 3D plot of the periodic orbits (without the asteroid) relative to the body-fixed frame of 433 Eros; (c) Distribution of six characteristic multipliers of the periodic orbit (the smallest periodic orbit among the seven periodic orbits plotted), the period is 8.656684h, the ratio of the period of the periodic orbit and the period of the asteroid is 1.64400724599; (d) Distribution of six characteristic multipliers of the periodic orbit (the biggest periodic orbit among the seven periodic orbits plotted) , the period is 6.684357h, the ratio of the period of the periodic orbit and the period of the asteroid is 1.26943881082.

Table 6. The initial positions and the initial velocities of periodic orbits in the body-fixed frame of 433 Eros (index 1 to 7 means the orbits from inner to outer)

| Periodic Orbits | Positions | Velocities | Periods |
|---|---|---|---|
| 1 | 0.226020373236<br>0.831549571341<br>0.000815731943 | 3.05135726080<br>-0.755105170473<br>-0.000787043149240 | 1.64400724599 |
| 2 | 0.162260294966<br>0.925373223566 | 3.69842543490<br>-0.620219330154 | 1.53615360513 |

|   | 0.000698149394 | -0.000317281286244 |   |
|---|---|---|---|
| 3 | 0.118572781769 | 4.31126220291 | 1.45298231634 |
|   | 1.01015364846 | -0.494283666047 |   |
|   | 0.000592999780 | -0.000142173246285 |   |
| 4 | 0.039313088475 | 4.90077984986 | 1.38965306382 |
|   | 1.09256772994 | -0.174726400201 |   |
|   | 0.000508929461 | -0.0462158783466 |   |
| 5 | 0.0877117956974 | 5.43432437124 | 1.34047716511 |
|   | 1.16503659385 | -0.405607013915 |   |
|   | 0.000440777741409 | -0.0000433255175181 |   |
| 6 | -0.0194496986212 | 5.97936660822 | 1.30132599619 |
|   | 1.24241338092 | 0.0923344130837 |   |
|   | 0.000386702840792 | -0.00000568727917105 |   |
| 7 | -0.01819579281 | 6.50081649913 | 1.26943881082 |
|   | 1.31630038108 | 0.0889931429895 |   |
|   | 0.000342011597241 | -0.00000361670859952 |   |

For the asteroid 101955 Bennu, the rotation period is 4.288h. For the asteroid 6489 Golevka, the rotation period is 6.026h.

Figure 11 shows a periodic orbit around asteroid 101955 Bennu, and Figure 12 shows a periodic orbit around asteroid 6489 Golevka. Both of the two periodic orbits are nearly circular, stable, and retrograde with zero inclination. The initial positions and velocities of these two periodic orbits in the periodic orbit family in the gravitational potential of the asteroids are presented in Table 7. Furthermore, the periodic orbit presented in Figure 12 has the period ratio 0.495486995181, which is near the 1:2 resonance of the period of the periodic orbit and the rotation period of the asteroid 6489 Golevka. Jiang et al. (2015c) found periodic orbits in the potential of asteroid 216 Kleopatra; the periodic orbits are nearly circular retrograde and stable with zero inclination; the asteroid 216 Kleopatra has seven equilibrium points. From Figure 9 to Figure 12, one can see that the periodic orbits which are nearly circular, retrograde and stable with zero inclination exist not only on the gravitational field of asteroid 216 Kleopatra, but also in the gravitational field of 433 Eros, 101955 Bennu and 6489 Golevka. This implies that the periodic orbits which are nearly circular retrograde and stable with zero inclination may widely exist in the potential of different gravitational field structure of minor celestial bodies.

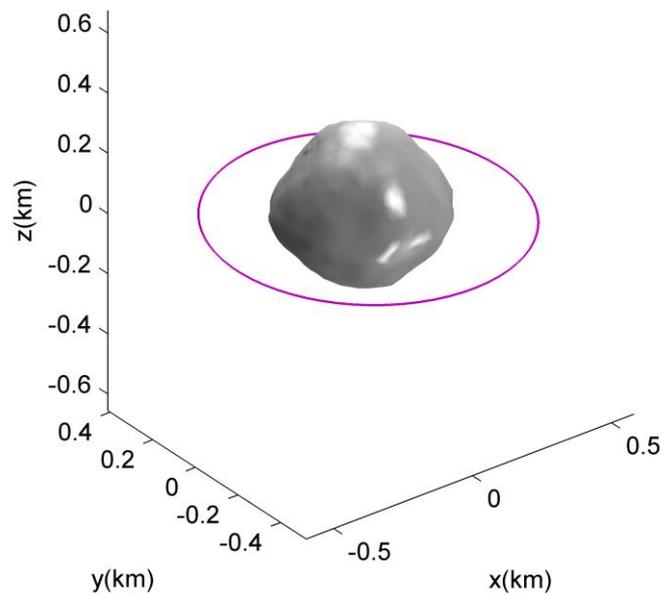

(a)

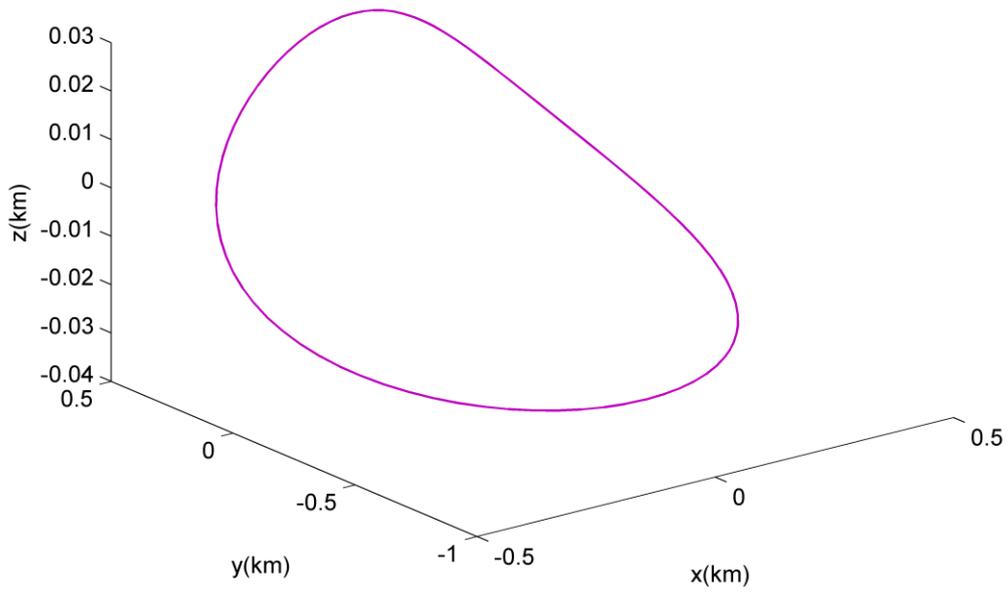

(b)

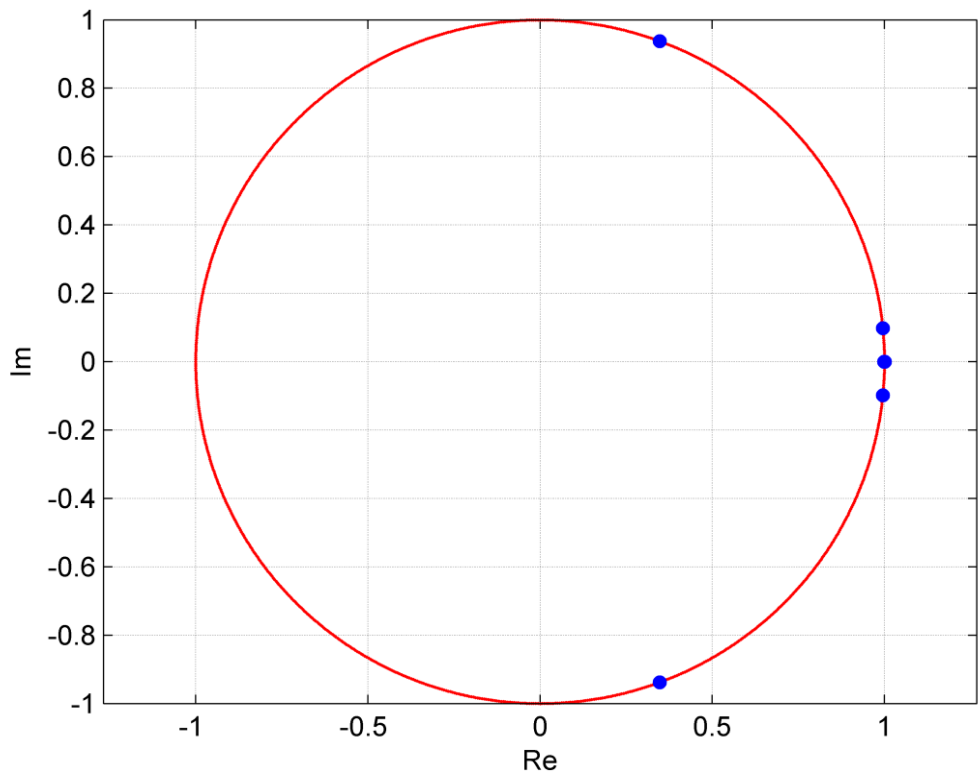

(c)

Figure 11. A nearly circular periodic orbit around asteroid 101955 Bennu which is retrograde with zero inclination, the period is 8.41525971074304h, and the ratio of the period of the periodic orbit and the period of the comet is 1.96251392508. (a) The 3D plot of the periodic orbit relative to the body-fixed frame of 101955 Bennu; (b) The 3D plot of the periodic orbit relative to the body-fixed frame of 101955 Bennu (without the asteroid); (c) The distribution of six characteristic multipliers of the periodic orbit.

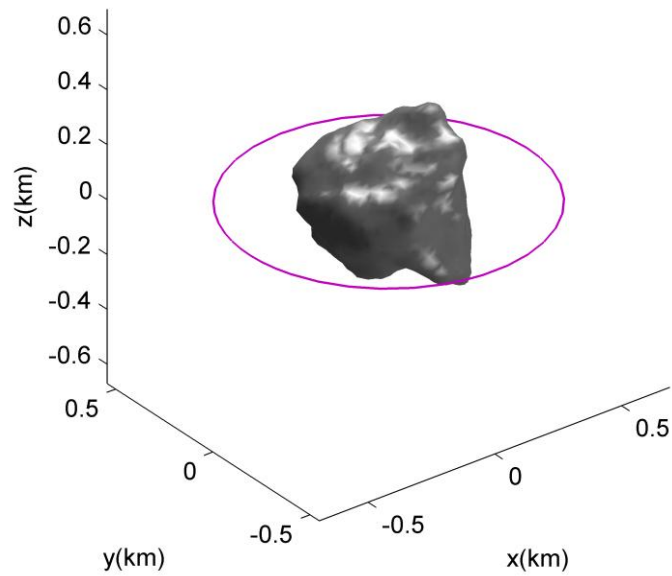

(a)

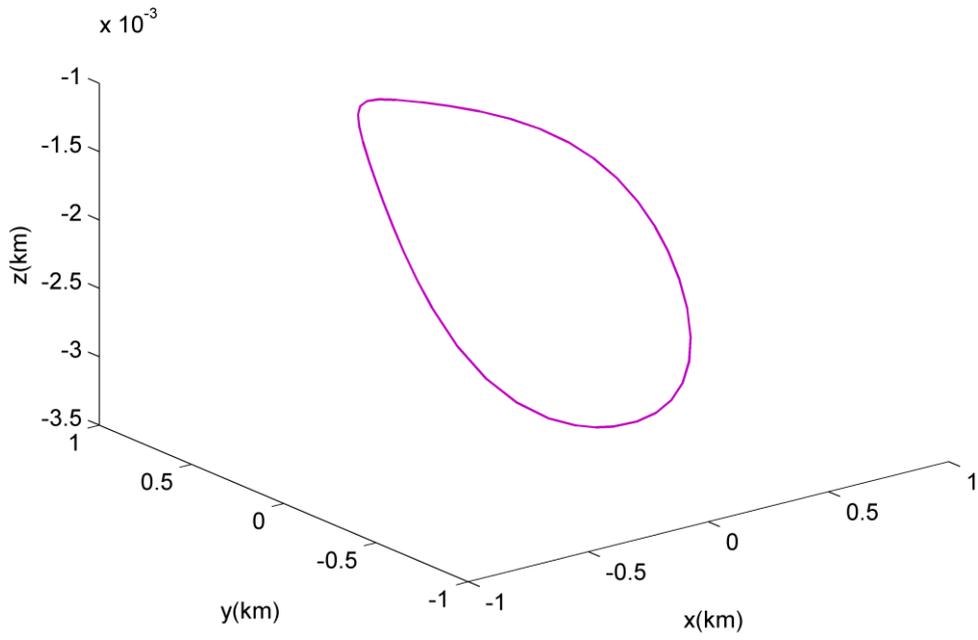

(b)

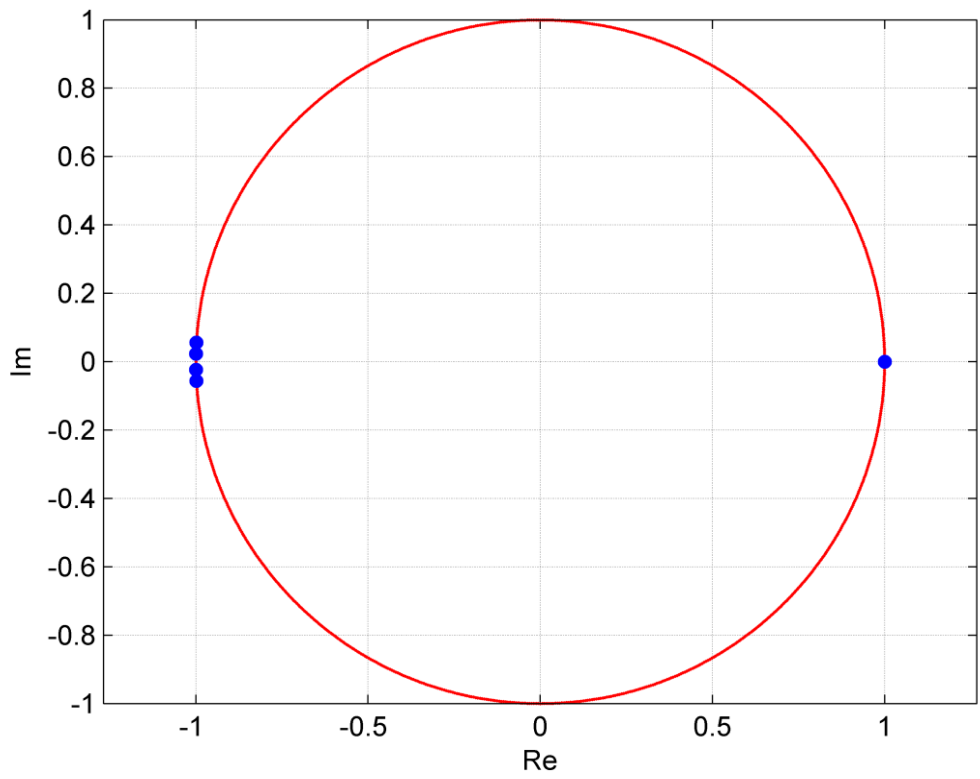

(c)

Figure 12. A nearly circular periodic orbit around asteroid 6489 Golevka which is retrograde with zero inclination, the period is 2.985804632960706h, and the ratio of the period of the periodic orbit and the period of the asteroid is 0.495486995181. (a) The 3D plot of the periodic orbit relative to the body-fixed frame of 6489 Golevka; (b) The 3D plot of the periodic orbit relative to the body-fixed frame of 6489 Golevka (without the asteroid); (c) The distribution of six characteristic multipliers of the periodic orbit.

Table 7. The initial positions and the initial velocities of periodic orbits in the body-fixed frame of 101955 Bennu and 6489 Golevka (periodic orbit 1 is the orbit around 101955 Bennu and periodic orbit 2 is the orbit around 6489 Golevka).

| Periodic Orbits | Positions | Velocities | Periods |
|---|---|---|---|
| 1 | 0.671591383153 | -2.01971118787 | 1.96251392508 |
|   | -0.586009389218 | -2.60968398375 |   |
|   | -0.0631586635050 | -0.0536311486152 |   |
| 2 | -0.802329098318 | 0.927694676293 | 0.495486995181 |
|   | 0.0683264606186 | 10.1907427252 |   |
|   | -0.00221645697675 | 0.0258563463006 |   |

**4.4 Resonance**

In this section, resonant periodic orbits means that the ratio of the period of the periodic orbit relative to the period of the asteroid is an integer. Russell [36] presented a 17:21 periodic orbit around asteroid 4 Vesta and several trajectories around highly irregular asteroid 4179 Toutatis. Jiang et al. [16] presented 4 resonant periodic orbits around asteroid 101955 Bennu, and one of them is stable. 101955 Bennu is the only one minor celestial body which has the maximum number of equilibrium points in the current study. There are totally 9 equilibrium points in the potential of asteroid 101955 Bennu, and 8 of them are outside the body.

Here we find resonant periodic orbits around other minor celestial bodies, including asteroid 243 Ida and 6489 Golevka. The length unit and time unit for motion around 243 Ida are the same them in Section 4.2. The length unit and time unit for motion around 6489 Golevka are the same them in Section 4.3.

Figure 13 presents a resonant periodic orbit around the primary of the binary asteroid 243 Ida. From Figure 13(b), one can see that the periodic orbit is stable, and has four characteristic multipliers equal 1. The initial positions and velocities of the periodic orbit in the gravitational potential of 243 Ida are presented in Table 8. The vectors are expressed in the asteroid's body-fixed frame. The periodic orbit is calculated using the Poincaré section; it has 8 intersections with the Poincaré section.

Figure 14 and 15 present two 2:1 resonant periodic orbits around the asteroid 6489 Golevka. From Figure 14(b) and 15(b), one can see that these two periodic orbits are stable. The periodic orbit presented in Figure 14 has two characteristic multipliers equal 1, while the periodic orbit presented in Figure 15 has four characteristic multipliers equal 1. The initial positions and velocities of the periodic orbits in the periodic orbit family in the gravitational potential of 6489 Golevka are presented in Table 8. The vectors are expressed in the asteroid's body-fixed frame. These two periodic orbits are calculated using the Poincaré section; both of them have 3 intersections with the Poincaré section.

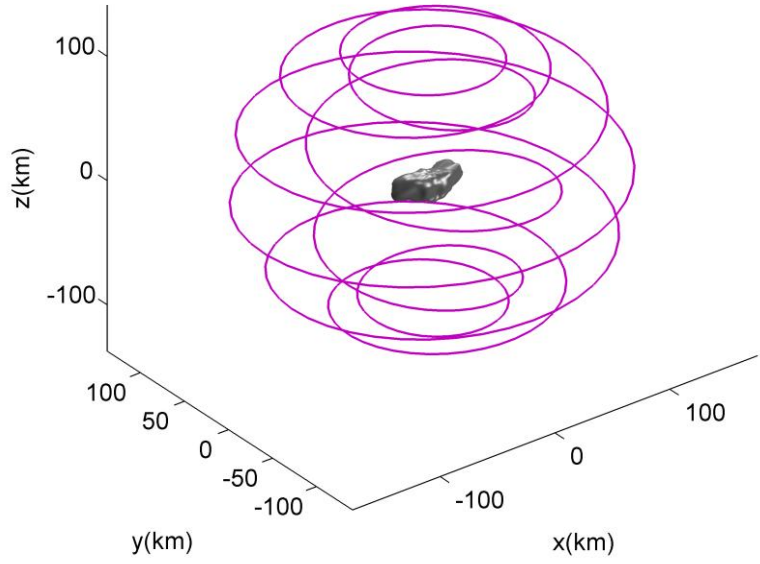

(a)

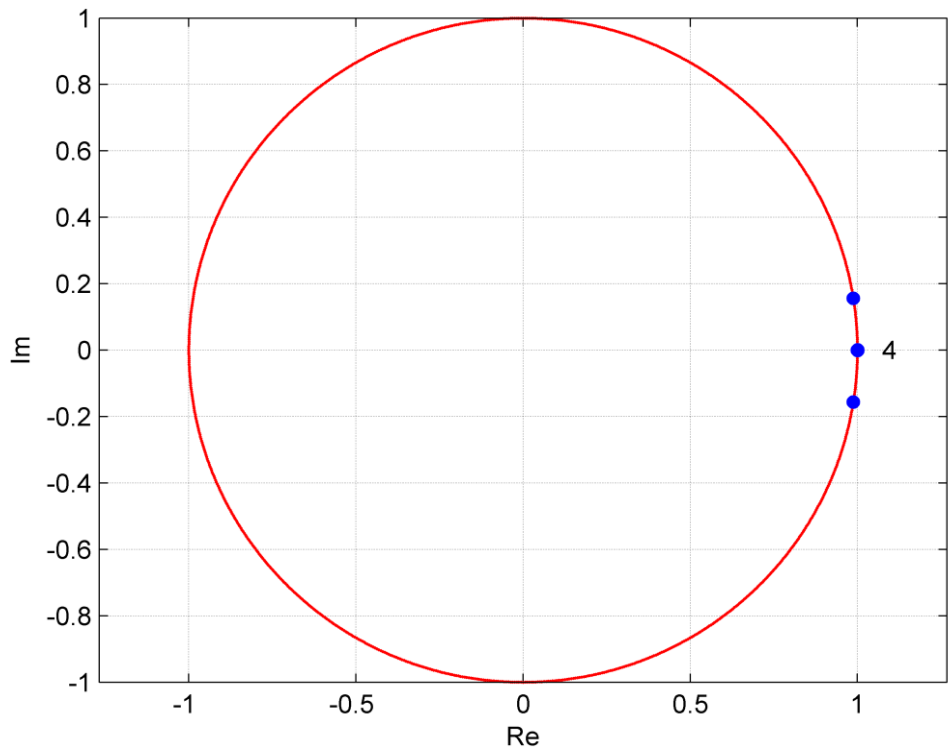

(b)

Figure 13. A resonant periodic orbit around asteroid 243 Ida, the ratio of the period of the periodic orbit and the period of the comet is 9.00525500454. (a) The 3D plot of the periodic orbit relative to the body-fixed frame of 243 Ida; (b) The distribution of six characteristic multipliers of the periodic orbit.

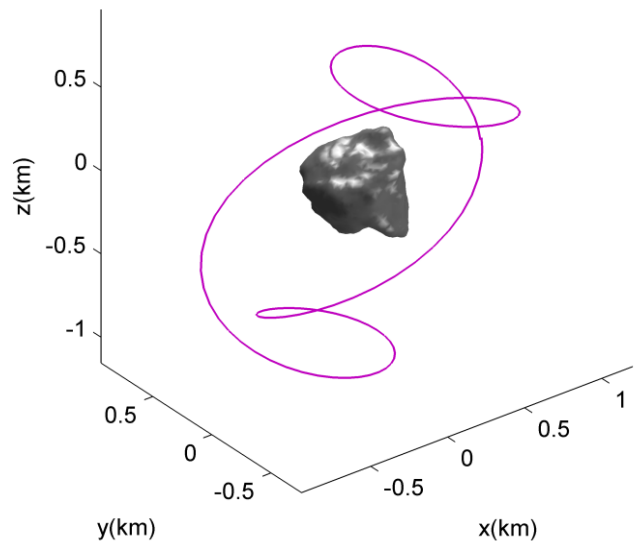

(a)

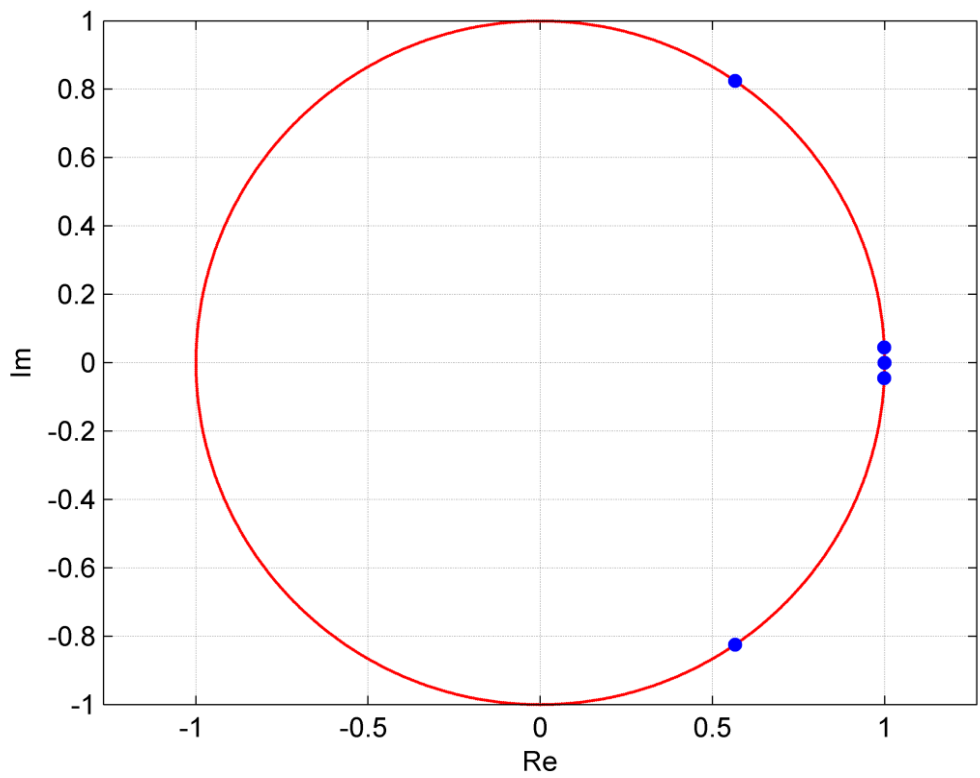

(b)

Figure 14. A resonant periodic orbit around asteroid 6489 Golevka, the ratio of the period of the periodic orbit and the period of the comet is 2.00558121907. (a) The 3D plot of the periodic orbit relative to the body-fixed frame of 6489 Golevka; (b) The distribution of six characteristic multipliers of the periodic orbit.

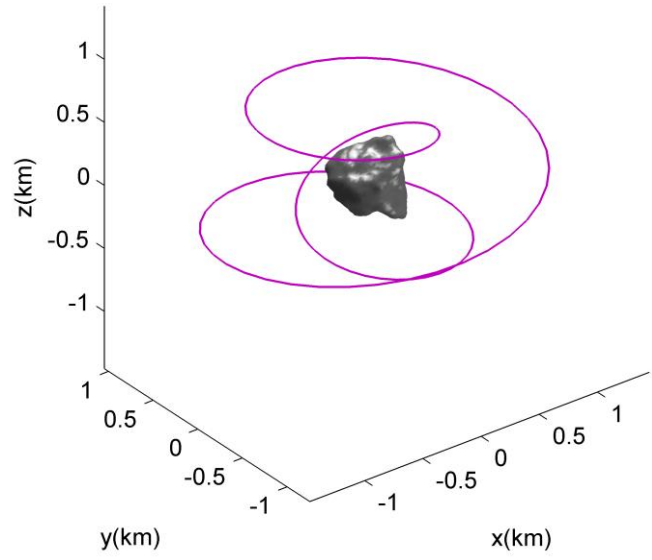

(a)

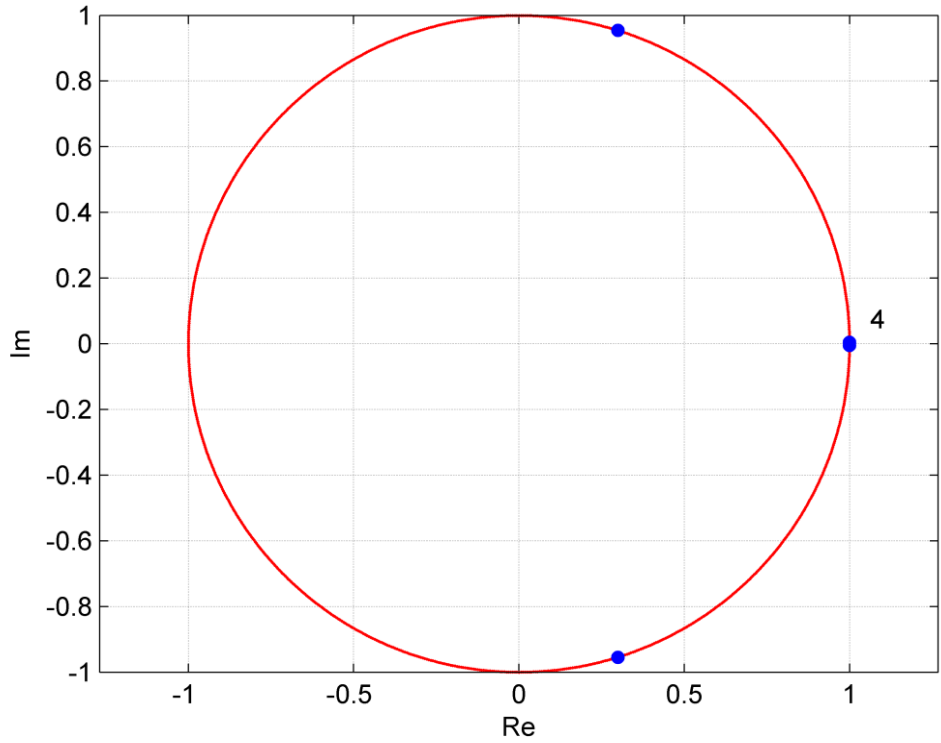

(b)

Figure 15. A resonant periodic orbit around asteroid 6489 Golevka, the ratio of the period of the periodic orbit and the period of the comet is 2.01369508451. (a) The 3D plot of the periodic orbit relative to the body-fixed frame of 6489 Golevka; (b) The distribution of six characteristic multipliers of the periodic orbit.

Table 8. The initial positions and the initial velocities of periodic orbits in the body-fixed frame of 243 Ida and 6489 Golevka (Periodic orbits 1, 2, and 3 are the initial values plotted in Figure 13, 14, and 15, respectively. Periodic orbits 1 is around 243 Ida, periodic orbit 2 and 3 are around 6489 Golevka)

| Periodic Orbits | Positions | Velocities | Periods | Resonant ratios |
|---|---|---|---|---|
| 1 | 0.897672329355<br>-1.29090224639<br>-0.0477816158615 | -8.73659445212<br>-6.14694020168<br>1.55083729969 | 9.00525500454 | 9:1 |
| 2 | 0.616839318105<br>-0.719651400484<br>0.535816234907 | -3.66314369527<br>-6.19748150670<br>-3.96827654611 | 2.00558121907 | 2:1 |
| 3 | 0.470335130163<br>0.555139854298<br>0.164844445024 | 6.89848549170<br>-6.28441207004<br>4.07158633235 | 2.01369508451 | 2:1 |

**4.5 Near-surface inclined periodic orbits**

The length unit and time unit for motion around the comet 1P/Halley are the same as them in Section 4.1. Figure 16 shows two different near-surface inclined periodic orbits, and these periodic orbits are non-resonant. From Figure 16(b), one can see that these two periodic orbits are really inclined relative to the xy plane (i.e. the equatorial plane of the comet). From Figure 16, one can also see that these two periodic orbits are near the surface of the body and stable. The initial positions and velocities of these two periodic orbits in the gravitational potential of 1P/Halley are presented in Table 9. The vectors are expressed in the comet's body-fixed frame. It is a remarkable fact that the ratio of the period of these two periodic orbits and the rotation period is smaller than 0.2.

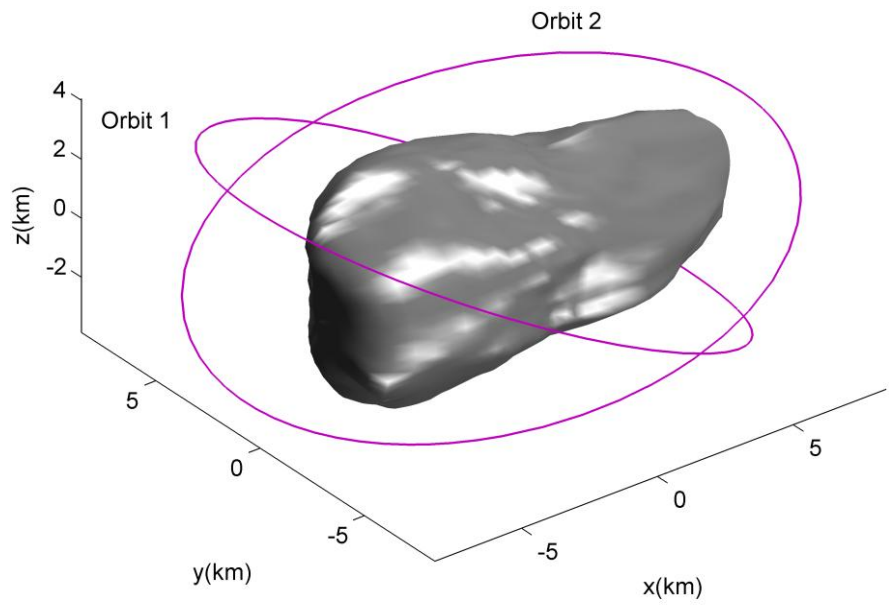

(a)

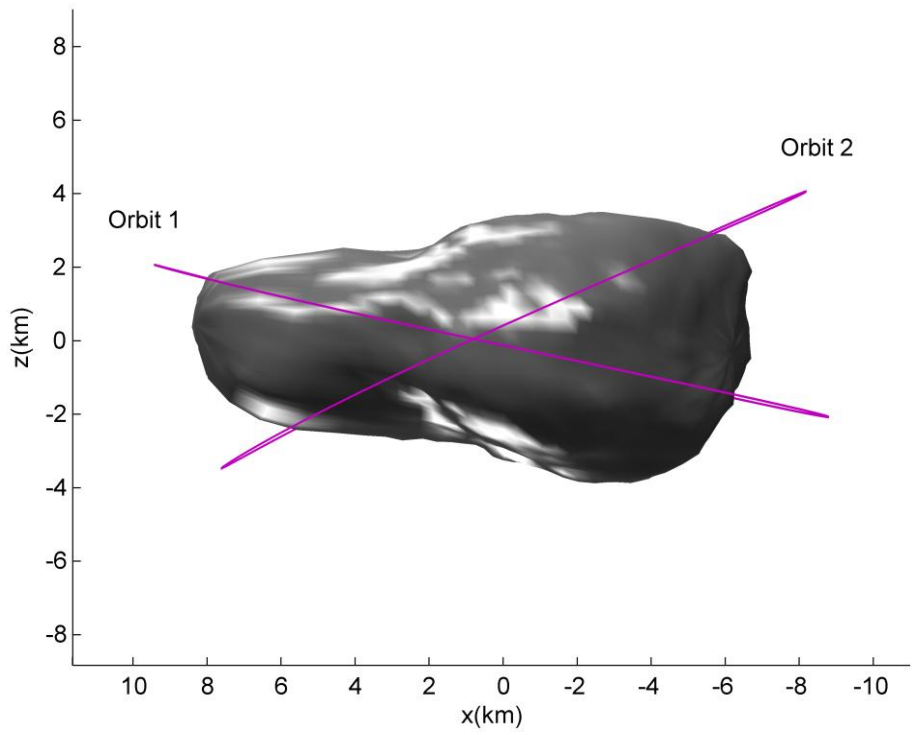

(b)

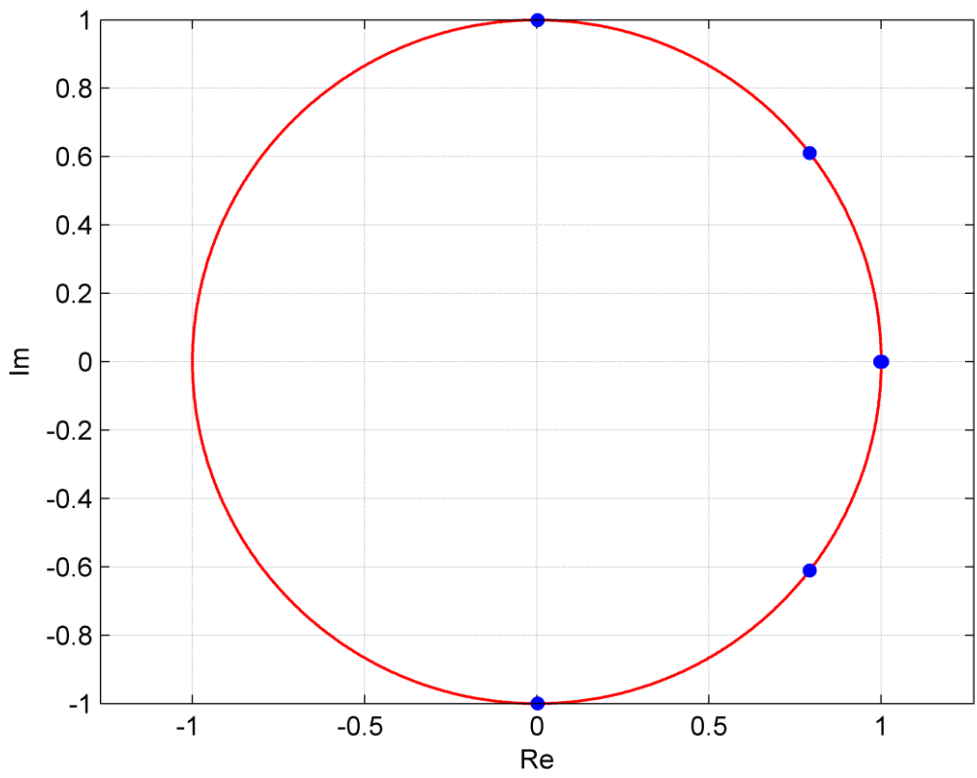

(c)

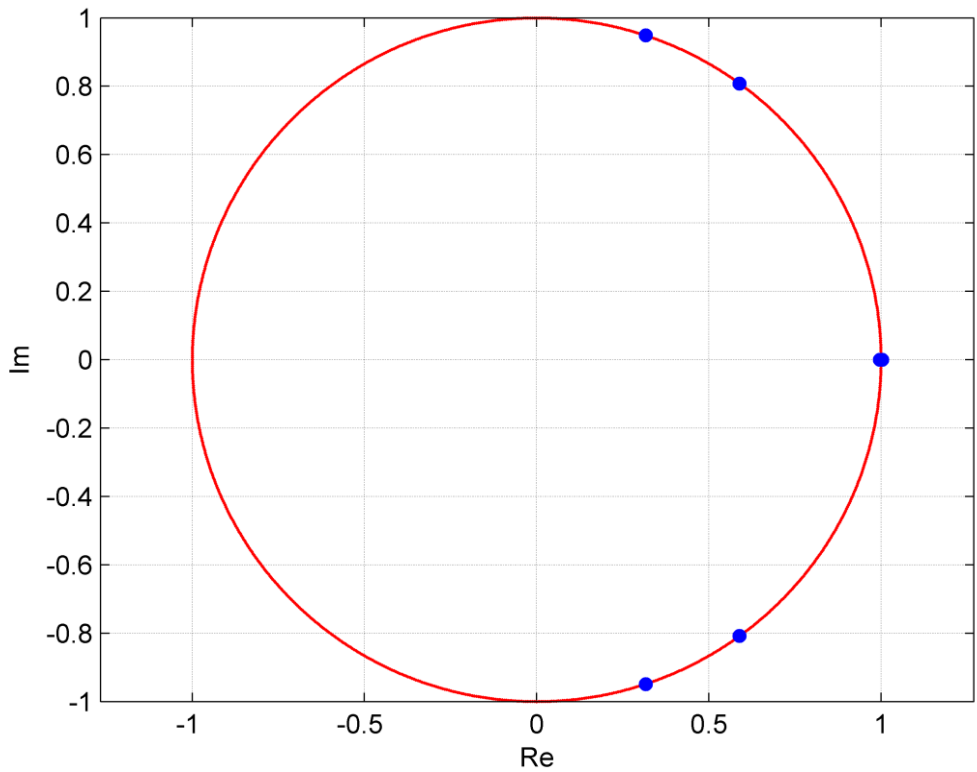

(d)

Figure 16. Two near-surface inclined periodic orbit around comet 1P/Halley. (a) The 3D plot of the periodic orbits relative to the body-fixed frame of 1P/Halley; (b) The periodic orbit relative to the body-fixed frame of 1P/Halley, viewed from +y axis; (c) Distribution of six characteristic multipliers of the periodic orbit 1; (d) Distribution of six characteristic multipliers of the periodic orbit 2.

Table 9. The initial positions and the initial velocities of periodic orbits in the body-fixed frame of 1P/ Halley

| Periodic Orbits | Positions | Velocities | Periods |
|---|---|---|---|
| 1 | 0.0629890965760<br>-0.587664950749<br>0.00132142402742 | -19.6763270413<br>-1.40740002652<br>-4.43845099506 | 0.180282795026 |
| 2 | -0.229339377692<br>0.517590660887<br>0.128585669518 | 16.6681557777<br>7.95258808390<br>-7.27379326767 | 0.174110986925 |

**4.6 Discussion of different kinds of stable periodic orbits**

From the above sections, we know that there exist several different kinds of stable periodic orbits around an irregular-shaped celestial body.

A) **Generated from the stable equilibrium points**: If the asteroid has stable equilibrium points outside the body, then there exist three families of stable periodic orbits around each of the stable equilibrium points [6, 24]. During the continuation, the periodic orbit families will change from stable to unstable (see Figure 17 for instance).

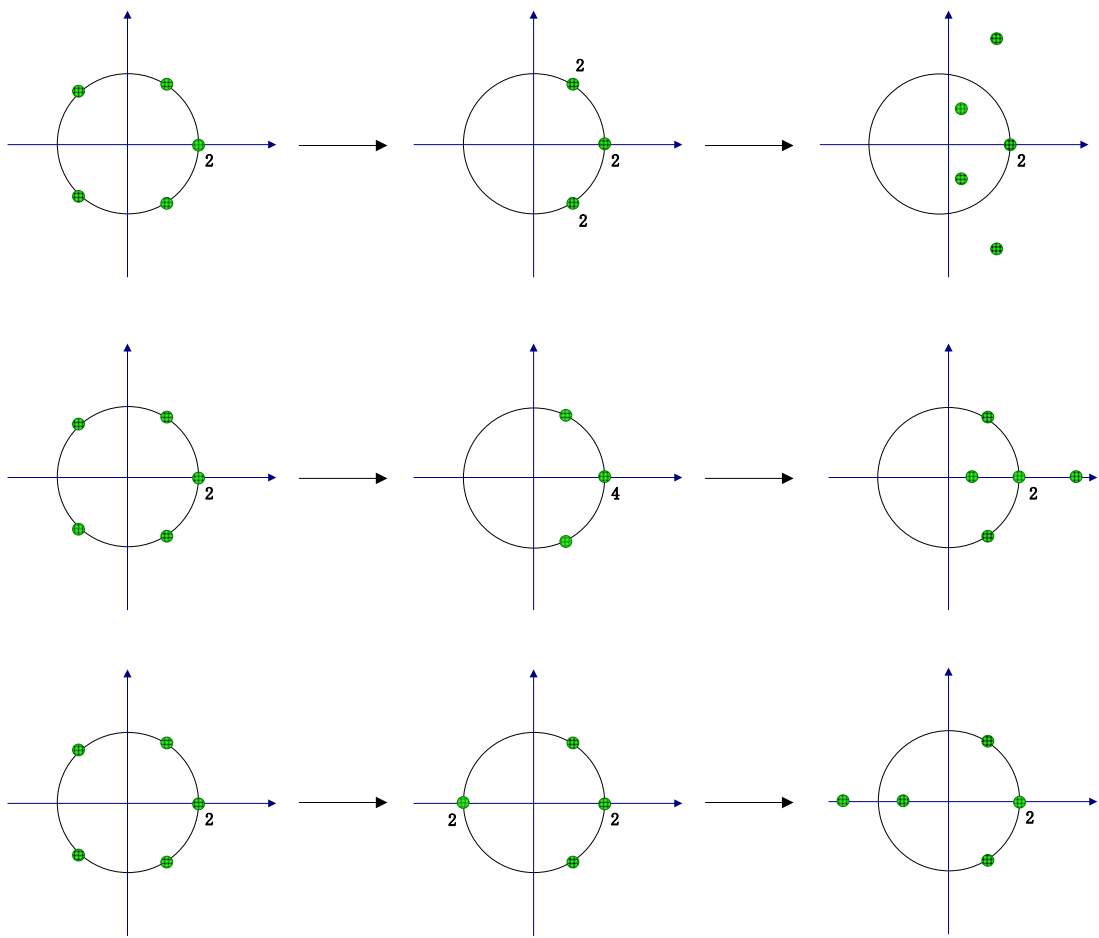

Figure 17. The examples of the stability variety of periodic orbit families generated

from the stable equilibrium points

B) **Continued from the unstable periodic orbits around the unstable equilibrium points**: If the asteroid has no stable equilibrium points outside the body, then there exists one family of unstable periodic orbits around the unstable equilibrium point if the topological classifications of the equilibrium point belong to Case 5 [6]; for Case 2 of the unstable equilibrium point, there exist two family of unstable periodic orbits around the equilibrium point. These unstable periodic orbits can be continued until the periodic orbits become stable.

C) **Retrograde and nearly circular periodic orbits with zero inclination**: There exist periodic orbit families which are nearly circular, and retrograde relative to the body-fixed frame of the asteroid [16, 18]. In addition, the inclinations of the periodic orbits relative to the asteroid's body-fixed frame are nearly equal to zero.

D) **Resonance**: There exist several resonant periodic orbit families which are stable [16].

E) **Near-surface inclined periodic orbits**: There exist several near-surface inclined periodic orbits around the asteroid, for which the distance between the periodic orbit and the mass center of the asteroid is smaller than the distance between the outside equilibrium point and the mass center of the asteroid.

Among these five different kinds of stable periodic orbits, A) has been found in Jiang et al. [6], C) and D) can be found in Jiang et al. [16]. B) and E) are found for the first time in this paper. However, the previous studies didn't study and classify different kinds of stable periodic orbits around minor celestial bodies. During the continuation of retrograde nearly circular, and stable periodic orbits with zero inclination around minor bodies, our results imply that the periodic orbits with the above characteristics can be found in the potential of different gravitational field structure of minor celestial bodies. In addition, there may be at least two periodic orbit families around one minor celestial body. Both of them are retrograde and nearly circular, and stable with zero inclination.

For the two near-surface inclined periodic orbits, it is notable that the ratio of the period of these two periodic orbits relative the rotation period is smaller than 0.2, which may be useful for us to search more kinds of near-surface inclined periodic orbits around other minor celestial bodies.

## 5. Conclusions

We analyze the stable periodic motions of spacecrafts in the gravitational field of minor celestial bodies. The grid search method is used to calculate the periodic orbits. We use the shapes, positions, inclinations, retrograde or not, topological classifications, continuation properties, section planes, as well as the resonance ratios to analyze the different kinds of stable periodic motions around asteroids and comets. Five kinds of stable periodic orbits are classified: A) Stable periodic orbits generated from the stable equilibrium points outside the minor celestial body; B) Stable periodic orbits continued from the unstable periodic orbits around the unstable equilibrium points; C) Periodic orbits which are retrograde and nearly circular with zero inclination; D) Resonant periodic orbits; E) near-surface inclined Stable periodic orbits.

Minor celestial bodies with different structure of gravitational fields are taken to calculate these stable periodic orbits, including asteroid 243 Ida, 433 Eros, 6489 Golevka, 101955 Bennu, and the comet 1P/Halley. The results of the stable periodic orbits around minor celestial bodies are useful for the mission design of the deep space exploration and the study the configurations and stabilities of small satellites in the large-size-ratio binary asteroids.

**Acknowledgements**

This research was supported by the State Key Laboratory of Astronautic Dynamics Foundation (No. 2016ADL0202).